\documentclass[twocolumn]{aastex63}

\usepackage{silence} 
\usepackage{times}
\usepackage{enumitem}
\usepackage{graphicx}
\usepackage{amsmath}
\usepackage{bm}
\usepackage{graphics}
\usepackage{url}
\usepackage{epsfig}
\usepackage{wrapfig}
\usepackage{ulem}
\usepackage{verbatim}
\usepackage{float}
\usepackage{lastpage}
\usepackage{longtable}

\usepackage{natbib}

\makeatletter
\define@key{Gin}{resolution}{\pdfimageresolution=#1\relax}
\makeatother

\providecommand{\apjl}{ApJ}

\providecommand{\apjs}{ApJS}
\providecommand{\aap}{A\&A}

\def\simlt{\mathrel{\hbox{\rlap{\hbox{\lower4pt\hbox{$\sim$}}}\hbox{$<$}}}}
\def\simgt{\mathrel{\hbox{\rlap{\hbox{\lower4pt\hbox{$\sim$}}}\hbox{$>$}}}}

\def\I{\,\textsc{i}}

\def\V{\,\textsc{v}}

\def\Mdot{$\,M_{\odot}$}

\def\arcsec{$^{\,\prime\prime}$}
\def\arcmin{$^{\,\prime}$}
\def\arcdeg{\degr}
 
\def\Ncandidate{189} % Candidates we consider
\def\Ninternal{7}    % Candidates from internal surveys
\def\Ngaia{0}        % Candidates ruled out by Gaia cut
\def\Nmpc{11}        % Candidates ruled out by MPC cut 
\def\Nclass{12}      % Candidates ruled out by spectroscopic classification
\def\Nvariable{2}    % Candidates ruled out by pre-merger variability cut

\def\Nstrong{114}
\def\Fstrong{41\%}
\def\Nspecz{31}
\def\Nphotoz{104}
\def\fphotoz{55\%}
\def\Nphot{28}

\shorttitle{GW190814 community search}
\shortauthors{Kilpatrick et~al.}

\begin{document}

\title{The Gravity Collective: A Search for the Electromagnetic Counterpart to the Neutron Star-Black Hole Merger GW190814}

\newcommand{\NU}{\affiliation{Center for Interdisciplinary Exploration and Research in Astrophysics (CIERA), Northwestern University, Evanston, IL 60208, USA}}
\newcommand{\fermi}{\affiliation{Fermi National Accelerator Laboratory, P. O. Box 500, Batavia, IL 60510, USA}}
\newcommand{\austinstate}{\affiliation{Austin Peay State University, Dept. Physics, Engineering and Astronomy, P.O. Box 4608, Clarksville, TN 37044, USA}}
\newcommand{\UCSC}{\affiliation{Department of Astronomy and Astrophysics, University of California, Santa Cruz, CA 95064, USA}}
\newcommand{\carnegie}{\affiliation{The Observatories of the Carnegie Institution for Science, 813 Santa Barbara St., Pasadena, CA 91101, USA}}
\newcommand{\kipac}{\affiliation{Kavli Institute for Particle Astrophysics \& Cosmology, P. O. Box 2450, Stanford University, Stanford, CA 94305, USA}}
\newcommand{\LBNL}{\affiliation{Lawrence Berkeley National Laboratory, 1 Cyclotron Road, MS 50B-4206, Berkeley, CA 94720-3411, USA}}
\newcommand{\benedictine}{\affiliation{Benedictine University, Department of Physics, 5700 College Road, Lisle, IL, 60532, USA}}
\newcommand{\ets}{\affiliation{Department of Physics and Astronomy, East Tennessee State University, Johnson City, TN 37614, USA}}
\newcommand{\CBPF}{\affiliation{Centro Brasileiro de Pesquisas F\'isicas, Rua Dr. Xavier Sigaud 150, CEP 22290-180, Rio de Janeiro, RJ, Brazil}}
\newcommand{\CEFET}{\affiliation{Centro Federal de Educa\c{c}\~ao Tecnol\'ogica Celso Suckow da Fonseca, Rodovia M\'ario Covas, lote J2, quadra J, CEP 23810-000,  Itagua\'i, RJ, Brazil}}
\newcommand{\ICAS}{\affiliation{International Center for Advanced Studies \& Instituto de Ciencias Físicas,  ECyT-UNSAM \& CONICET, 1650, Buenos Aires, Argentina}}
\newcommand{\NATU}{\affiliation{NAT-Universidade Cruzeiro do Sul / Universidade Cidade de S{\~a}o Paulo, Rua Galv{\~a}o Bueno, 868, 01506-000, S{\~a}o Paulo, SP, Brazil}}
\newcommand{\IITH}{\affiliation{Indian Institute of Technology, Hyderabad, Kandi Telangana 502285 India}}
\newcommand{\UCBerkeley}{\affiliation{Department of Astronomy, University of California, Berkeley, CA 94720-3411, USA}}
\newcommand{\Miller}{\affiliation{Miller Institute for Basic Research in Science, University of California, Berkeley, CA 94720, USA}}
\newcommand{\IAPUCC}{\affiliation{Instituto de Astrof\'isica, Pontificia Universidad Cat\'olica de Chile, Casilla 306, Santiago 22, Chile}}
\newcommand{\millennium}{\affiliation{Millennium Institute of Astrophysics (MAS), Nuncio Monse$\tilde{n}$or S\'otero Sanz 100, Providencia, Santiago, Chile}}
\newcommand{\carnegieLCO}{\affiliation{Carnegie Observatories, Las Campanas Observatory, Casilla 601, La Serena, Chile}}
\newcommand{\CCNE}{\affiliation{Departamento de F\'isica, Centro de Ci\^encias Naturais e Exatas, Universidade Federal de Santa Maria, 97105-900, Santa Maria, RS, Brazil}}
\newcommand{\UDP}{\affiliation{N\'ucleo de Astronom\'ia, Universidad Diego Portales, Av. Ejército 441, Santiago, Chile}}
\newcommand{\Michigan}{\affiliation{Department of Physics, University of Michigan, Ann Arbor, MI 48109, USA}}
\newcommand{\UWMadison}{\affiliation{Physics Department, University of Wisconsin-Madison, Madison, WI 53706, USA}}
\newcommand{\Steward}{\affiliation{Steward Observatory, University of Arizona, 933 North Cherry Avenue, Tucson, AZ 85721-0065, USA}}
\newcommand{\TelAviv}{\affiliation{The School of Physics and Astronomy, Tel Aviv University, Tel Aviv 69978, Israel}}
\newcommand{\CIFAR}{\affiliation{CIFAR Azrieli Global Scholars program, CIFAR, Toronto, Canada}}
\newcommand{\MtStromlo}{\affiliation{Mt Stromlo Observatory, The Research School of Astronomy and Astrophysics, Australian National University, ACT 2601, Australia}}
\newcommand{\NCPAS}{\affiliation{National Centre for the Public Awareness of Science, Australian National University, Canberra, ACT 2611, Australia}}
\newcommand{\ARC}{\affiliation{The ARC Centre of Excellence for All-Sky Astrophysics in 3 Dimensions (ASTRO 3D), Australia}}
\newcommand{\ColumbiaAL}{\affiliation{Columbia Astrophysics Laboratory, Columbia University, New York, NY 10027, USA}}
\newcommand{\flatiron}{\affiliation{Center for Computational Astrophysics, Flatiron Institute, 162 W. 5th Avenue, New York, NY 10011, USA}}
\newcommand{\IU}{\affiliation{Department of Astronomy, Indiana University, 727 E. Third St., Bloomington, IN 47405}}
\newcommand{\Racah}{\affiliation{Racah Institute for Physics, The Hebrew University, Jerusalem Israel 91904}}
\newcommand{\MPIA}{\affiliation{Max-Planck-Institut fur Astrophysik, Karl-Schwarzschild-Str 1, D-85748 Garching bei M\"unchen, Germany}}
\newcommand{\UCSB}{\affiliation{Department of Physics, University of California, Santa Barbara, CA 93106-9530, USA}}
\newcommand{\LCO}{\affiliation{Las Cumbres Observatory, 6740 Cortona Dr, Suite 102, Goleta, CA 93117-5575, USA}}
\newcommand{\efermi}{\affiliation{Enrico Fermi Institute, Department of Physics, Department of Astronomy and Astrophysics}}
\newcommand{\kavlicosmo}{\affiliation{Enrico Fermi Institute, Department of Physics, Department of Astronomy and Astrophysics,\\and Kavli Institute for Cosmological Physics, University of Chicago, Chicago, IL 60637, USA}}
\newcommand{\jordell}{\affiliation{Jodrell Bank Centre for Astrophysics, University of Manchester, Oxford Road, Manchester, UK}}
\newcommand{\UCD}{\affiliation{Department of Physics and Astronomy, University of California, Davis, CA, 95616}}
\newcommand{\IFT}{\affiliation{Instituto de F\'isica Te\'orica UAM/CSIC, Universidad Aut\'onoma de Madrid, 28049 Madrid, Spain}}
\newcommand{\CAS}{\affiliation{Centre for Astrophysics and Supercomputing, Swinburne University of Technology, PO Box 218, H29, Hawthorn, VIC, 3122, Australia}}
\newcommand{\OzGrav}{\affiliation{Australian Research Council Centre of Excellence for Gravitational Wave Discovery, Swinburne University of Technology, Hawthorn, VIC, 3122, Australia}}
\newcommand{\UAB}{\affiliation{Departamento de Ciencias Fisicas, Universidad Andres Bello, Avda. Republica 252, Santiago, Chile}}
\newcommand{\UQB}{\affiliation{School of Mathematics and Physics, University of Queensland, Brisbane, QLD, 4072, Australia}}
\newcommand{\weizmann}{\affiliation{Department of Particle Physics and Astrophysics, Weizmann Institute of Science, Rehovot, 7610001, Israel}}
\newcommand{\UChicago}{\affiliation{Department of Astronomy and Astrophysics, University of Chicago, Chicago, IL 60637, USA}}
\newcommand{\atacama}{\affiliation{Instituto de Astronom\'{\i}a y Ciencias Planetarias, Universidad de Atacama, Copayapu 485, Copiap\'o, Chile}}
\newcommand{\Thacher}{\affiliation{Thacher Observatory, Thacher School, 5025 Thacher Rd. Ojai, CA 93023, USA}}
\newcommand{\Hawaii}{\affiliation{Institute for Astronomy, University of Hawaii, 2680 Woodlawn Drive, Honolulu, HI 96822, USA}}
\newcommand{\camino}{\affiliation{Departamento de Astronom\'ia, Universidad de Chile, Camino El Observatorio 1515, Las Condes, Santiago, Chile}}
\newcommand{\JHU}{\affiliation{Department of Physics and Astronomy, Johns Hopkins University, 3400 North Charles Street, Baltimore, MD 21218, USA}}
\newcommand{\STScI}{\affiliation{Space Telescope Science Institute, 3700 San Martin Drive, Baltimore, MD 21218, USA}}
\newcommand{\mcgill}{\affiliation{Department of Physics, McGill University, 3600 University Street, Montr\'eal, QC H3A 2T8, Canada}}
\newcommand{\mcgillinst}{\affiliation{McGill Space Institute, McGill University, 3550 University Street, Montr\'eal, QC H3A 2A7, Canada}}
\newcommand{\cdita}{\affiliation{Center for Data Intensive and Time Domain Astronomy, Department of Physics and Astronomy, Michigan State University, East Lansing, MI 48824, USA}}
\newcommand{\cfa}{\affiliation{Center for Astrophysics, Harvard \& Smithsonian, 60 Garden St, Cambridge, MA 02138, USA}}
\newcommand{\Einstein}{\altaffiliation{NASA Einstein Fellow}}
\author[0000-0002-5740-7747]{Charles D. Kilpatrick}
\NU

\author[0000-0003-4263-2228]{David A. Coulter}
\UCSC

\author[0000-0001-7090-4898]{Iair Arcavi}
\TelAviv\CIFAR

\author[0000-0001-5955-2502]{Thomas G. Brink}
\UCBerkeley

\author[0000-0001-9494-179X]{Georgios Dimitriadis}
\UCSC

\author{Alexei V. Filippenko}
\UCBerkeley
\Miller

\author[0000-0002-2445-5275]{Ryan J. Foley}
\UCSC

\author[0000-0003-4253-656X]{D. Andrew Howell}
\UCSB\LCO

\author{David O. Jones}
\UCSC\Einstein

\author[0000-0003-2206-2651]{Martin Makler}
\ICAS\CBPF

\author[0000-0001-6806-0673]{Anthony L. Piro}
\carnegie

\author[0000-0002-7559-315X]{C\'esar Rojas-Bravo}
\UCSC

\author[0000-0003-4102-380X]{David J. Sand}
\Steward

\author[0000-0002-9486-818X]{Jonathan J. Swift}
\Thacher

\author{Douglas Tucker}
\fermi

\author{WeiKang Zheng}
\UCBerkeley

\author{Sahar S. Allam}
\fermi

\author[0000-0002-0609-3987]{James T. Annis}
\fermi

\author{Juanita Antilen}
\camino

\author[0000-0002-6119-5353]{Tristan G. Bachmann}
\UChicago

\author[0000-0002-7777-216X]{Joshua S. Bloom}
\LBNL

\author[0000-0003-4383-2969]{Clecio R. Bom}
\CBPF\CEFET

\author[0000-0002-4924-444X]{K. Azalee Bostroem}
\UCD

\author[0000-0001-5201-8374]{Dillon Brout}
\cfa\Einstein

\author[0000-0003-0035-6659]{Jamison Burke}
\UCSB\LCO

\author[0000-0003-2789-3817]{Robert E. Butler}
\IU

\author{Melissa Butner}
\ets

\author[0000-0002-3829-9920]{Abdo Campillay}
\carnegieLCO

\author{Karoli E. Clever}
\UCSC

\author[0000-0003-1949-7638]{Christopher J. Conselice}
\jordell

\author[0000-0001-5703-2108]{Jeff Cooke}
\CAS
\OzGrav

\author{Kristen C. Dage}
\mcgill\mcgillinst\cdita

\author{Reinaldo R. de Carvalho}
\NATU

\author{Thomas de Jaeger}
\UCBerkeley
\Hawaii

\author[0000-0002-0466-3288]{Shantanu Desai}
\IITH

\author{Alyssa Garcia}
\Michigan

\author[0000-0002-9370-8360]{Juan Garcia-Bellido} 
\IFT

\author[0000-0003-2524-5154]{Mandeep S. S. Gill}
\kipac

\author{Nachiket Girish}
\UCBerkeley

\author[0000-0002-0430-7793]{Na'ama Hallakoun}
\weizmann\TelAviv

\author[0000-0001-6718-2978]{Kenneth Herner}
\fermi

\author[0000-0002-1125-9187]{Daichi Hiramatsu}
\UCSB\LCO

\author[0000-0002-0175-5064]{Daniel E. Holz}
\kavlicosmo

\author{Grace Huber}
\Thacher

\author{Adam M. Kawash}
\cdita

\author[0000-0001-5807-7893]{Curtis McCully}
\UCSB\LCO

\author{Sophia A. Medallon}
\UCSC

\author[0000-0002-4670-7509]{Brian D. Metzger}
\ColumbiaAL\flatiron

\author{Shaunak Modak}
\UCBerkeley

\author[0000-0002-7016-5471]{Robert Morgan}
\UWMadison

\author{Ricardo R. Mu\~noz}
\camino

\author{Nahir Mu\~noz-Elgueta}
\MPIA

\author[0000-0002-8342-3804]{Yukei S. Murakami}
\UCBerkeley

\author[0000-0002-5115-6377]{Felipe Olivares E.}
\atacama

\author[0000-0002-6011-0530]{Antonella Palmese}
\fermi

\author{Kishore Patra}
\UCBerkeley

\author[0000-0002-7131-7684]{Maria E. S. Pereira}
\Michigan

\author{Thallis L. Pessi}
\CCNE\UDP

\author{J. Pineda-Garcia}
\UAB

\author{Jonathan Quirola-V\'asquez}
\IAPUCC\millennium

\author{Enrico Ramirez-Ruiz}
\UCSC

\author[0000-0003-0880-5738]{Sandro Barboza Rembold}
\CCNE

\author[0000-0002-4410-5387]{Armin Rest}
\JHU\STScI

\author[0000-0001-8651-8772]{\'{O}smar Rodr\'{i}guez}
\TelAviv

\author[0000-0003-3402-6164]{Luidhy Santana-Silva}
\NATU

\author{Nora F. Sherman}
\Michigan

\author{Matthew R. Siebert}
\UCSC

\author{Carli Smith}
\UCSC

\author[0000-0002-6261-4601]{J. Allyn Smith}
\austinstate

\author[0000-0001-6082-8529]{Marcelle Soares-Santos}
\Michigan

\author{Holland Stacey}
\Thacher

\author{Benjamin E. Stahl}
\UCBerkeley

\author{Jay Strader}
\cdita

\author{Erika Strasburger}
\UCBerkeley

\author{James Sunseri}
\UCBerkeley

\author{Samaporn Tinyanont}
\UCSC

\author[0000-0002-4283-5159]{Brad E. Tucker}
\MtStromlo\NCPAS\ARC

\author{Natalie Ulloa}
\carnegieLCO

\author{Stefano Valenti}
\UCD

\author{Sergiy Vasylyev}
\UCBerkeley

\author[0000-0001-8653-7738]{Matthew P. Wiesner}
\benedictine

\author[0000-0002-9955-8797]{Keto D. Zhang}
\UCBerkeley

\begin{abstract}

We present optical follow-up imaging obtained with the Katzman Automatic Imaging Telescope, Las Cumbres Observatory Global Telescope Network, Nickel Telescope, Swope Telescope, and Thacher Telescope of the LIGO/Virgo gravitational wave (GW) signal from the neutron star--black hole (NSBH) merger GW190814.  We searched the GW190814 localization region (19~deg$^{2}$ for the 90th percentile best localization), covering a total of 51~deg$^{2}$ and 94.6\% of the two-dimensional localization region. Analyzing the properties of \Ncandidate\ transients that we consider as candidate counterparts to the NSBH merger, including their localizations, discovery times from merger, optical spectra, likely host-galaxy redshifts, and photometric evolution, we conclude that none of these objects are likely to be associated with GW190814.  Based on this finding, we consider the likely optical properties of an electromagnetic counterpart to GW190814, including possible kilonovae and short gamma-ray burst afterglows.  Using the joint limits from our follow-up imaging, we conclude that a counterpart with an $r$-band decline rate of 0.68~mag~day$^{-1}$, similar to the kilonova AT~2017gfo, could peak at an absolute magnitude of at most $-17.8$~mag (50\% confidence).  Our data are not constraining for ``red'' kilonovae and rule out ``blue'' kilonovae with $M>0.5~M_{\odot}$ (30\% confidence).  We strongly rule out all known types of short gamma-ray burst afterglows with viewing angles $<${}17$^{\circ}$ assuming an initial jet opening angle of $\sim5.2^{\circ}$ and explosion energies and circumburst densities similar to afterglows explored in the literature.  Finally, we explore the possibility that GW190814 merged in the disk of an active galactic nucleus, of which we find four in the localization region, but we do not find any candidate counterparts among these sources.

\end{abstract}


%%%%%%%%%%%%%%%%%%%
%%  Introduction  %
%%%%%%%%%%%%%%%%%%%

\section{Introduction}\label{sec:intro}

Neutron star (NS) and black hole (BH) mergers are among the strongest gravitational wave (GW) sources from 10 to 10,000~Hz \citep{Press72,Thorne97} and the primary astrophysical sources detected by the Laser Interferometer Gravitational Wave Observatory (LIGO) and Virgo collaboration \citep[LVC;][]{Abbott:AdvLIGO,Abbott17:0814}.  Although electromagnetic (EM) follow-up observations of these events began with the first detection of a binary black hole (BBH) merger by LIGO \citep[][]{Abbott16:gw}, it was not until the discovery of the binary neutron star merger (BNS) GW170817 that EM and GW emission was observed from the same source \citep{Abbott17}.  GW170817 was accompanied by a prompt, short gamma-ray burst viewed off-axis \citep[sGRB;][]{Abbott17:grb,Savchenko17}\footnote{Although \citet{Kasliwal17} argue this event was much weaker than sGRBs viewed at high redshift and likely the result of a shock breakout.} and later a kilonova called AT~2017gfo\footnote{Also called SSS17a, DLT17ck, and PS17egl.} discovered at optical wavelengths \citep{Coulter17}.  Follow-up observations of this event spanned the EM spectrum, and combined with the GW data these observations enabled unique insight into the nature of its ejecta \citep[e.g.,][]{Arcavi17,Cowperthwaite17,Drout17,Kasliwal17,Kilpatrick17,Smartt17}, the engines that power sGRBs \citep{Abbott17:grb,Savchenko17,Fong17,2020arXiv200712245M}, and the NS equation of state \citep{Abbott17:radii,Radice18}.

The precise localization of GW170817 required coordination between the LVC and optical search teams \citep{Abbott17:detection,Coulter17}.  Critically, all three LVC detectors contributed to the localization of GW170817 and the distance to this event was only $\sim40$~Mpc from the initial LVC analysis \citep[][]{Abbott17:detection}.  This enabled a search of a relatively small volume of space that targeted galaxies in highly complete catalogs.  Indeed, the greatest limiting factors in the speed with which AT~2017gfo was identified were the timescale required to generate accurate localization maps and the positioning of telescopes across the globe \citep{Abbott17}. 

The same strategy has been less practical for all of the high-confidence NS mergers reported during LVC Observing Run 3 \citep[O3; including GW190425, S190426c, GW190814, S190910d, S190910h, S190923y, S190930t, S191205ah, S191213g, and S200213t in][]{Andreoni19,Andreoni19b,Coughlin19,Dobie19,Goldstein19b,Gomez19,Hosseinzadeh19,Lundquist19,Ackley20,Antier20,Coughlin20b,Morgan20,Paterson20,Pozanenko20,Thakur20,Vieira20,Watson20,Alexander21,deWet21}.  With greatly increased detector sensitivity and a higher rate of events detected at larger distances than GW170817, all LVC O3 events classified as NS mergers were less precisely localized than GW170817 with one exception.  Given the rapid decline rates expected for EM counterparts \citep{2011ApJ...736L..21R,Kasen14}, it is unlikely that a counterpart would be detected.  This is true even in the most optimistic counterpart models \citep{Andreoni19,Goldstein19b,Hosseinzadeh19,Morgan20,Thakur20,Alexander21,deWet21}, but especially after folding in realistic assumptions about the physical properties of NS mergers implied by GW data as in the case of GW190425 \citep{Foley20}.  Although some observations rule out AT~2017gfo-like counterparts over a large fraction of the localization regions of O3 events \citep[e.g.,][]{Coughlin20b,Goldstein19b,Morgan20}, the total ejecta mass, composition, and merger properties of most NS mergers remain almost entirely unconstrained.

The need for better constraints is most pressing for GW events from neutron star--black hole (NSBH) and black hole--black hole (BBH) mergers where no viable EM counterparts have been confirmed \citep[whereas GW170817 is widely considered to be the result of a BNS merger;][]{Abbott17,Kilpatrick17}.  Unlike BNS mergers where disruption of both NS components and some ejecta are guaranteed \citep{Li98,Shibata06,Metzger10,2011ApJ...736L..21R}, in a NSBH merger it is possible that the NS will eject no mass before it is entirely accreted by the BH.  This is because the NS must be tidally shredded before it reaches the innermost stable circular orbit to produce ejecta, and whether this happens depends on the total mass of the system, the mass ratio, spins, and the radius (and thus the equation of state) of the NS \citep{Faber06,2007NJPh....9...17L,Ferrari10}.  Thus, for each new NSBH, the LVC infers the probability that mass remains outside the merger based on numerical-relativity simulations \citep[i.e., {\tt HasRemnant} in][]{GCN25324,GCN25333}.  Values reported by the LVC use an optimistic (stiff) equation of state, which maximizes the value of {\tt HasRemnant} given constraints on the masses and spins of the merger components \citep{Abbott09}.  {\tt HasRemnant} is often interpreted as the likelihood of seeing an EM counterpart similar to a kilonova or sGRB \citep[][]{Kasen17,Troja17}.  Beyond comparisons to relatively simple systems such as the GW170817 merger where ejecta are guaranteed, the validity of this estimate and the nature of EM emission from NSBH mergers have yet to be verified, and thus all {\tt HasRemnant} estimates are subject to significant systematic uncertainties.

In the middle of O3 and roughly 2~yr after reporting GW170817, the LVC detected GW signal GW190814 on 14 Aug.\ 2019 at 21:11:16 UT \citep{GCN25324,GCN25333,Abbott20}.  Detailed analysis indicated that this signal had high significance, implying a relatively close event with a low probability of a ``false alarm'' \citep[with a rate of one per 10$^{11}$ Hubble times that a spurious signal from correlated noise or detector ``glitches'' could produce GW190814;][]{Singer16}.  GW190814 was classified as a NSBH with high significance \citep[$>$99\%; ][]{Abbott20}.  In the final analysis of the GW190814 strain, the best-fitting template to the GW strain signal being the merger of 2.59$\pm$0.08~$M_{\odot}$ and 23.2$\substack{+1.0\\-0.9}$~$M_{\odot}$ components at 235$\substack{+40\\-45}$~Mpc \citep{Abbott20}.  

In addition to being the best-localized and one of the closest GW events to date, GW190814 had one of the most extreme mass ratios of all GW events detected during the LVC's first three observing runs.  However, this analysis assumes that 2.59~$M_{\odot}$ NSs exist in nature, which may not be the case if the maximum-mass NS is below this threshold \citep{Fattoyev20,Tan20,Kanakis-Pegios21,Godzieba21,Wu21}.  The search for EM counterparts therefore provides unique insight into the nature of this threshold; detection of an EM counterpart from an NSBH merger with a massive secondary would imply that the maximum NS mass is at least as massive. Although analyses that classify GW190814 as a NSBH merger imply that the secondary is the most massive known NS, with significant implications for the compact binary population and their formation channels \citep{O3a}, it remains possible that GW190814 was a BBH system, incidentally making the secondary the least massive known BH.

The localization for GW190814 was rapidly refined to a localization region with size $\sim 38$~deg$^{2}$ (90th percentile) on 14 Aug.\ 2019 and centered approximately at $\alpha=24.6^{\circ}, \delta=-24.8^{\circ}$ (J2000), although the final localization map presented by \citet{Abbott20} had a 90th percentile area of 19~deg$^{2}$ (\autoref{fig:localization}).  Initial estimates of the {\tt HasRemnant} statistic by the LVC was $<1$\% \citep{GCN25324}, which is consistent with expectations that a 23.2~$M_{\odot}$+2.59~$M_{\odot}$ system would produce no ejecta even if the BH was maximally rotating.  Significant EM follow-up observations of this event were triggered by several groups \citep[][]{Andreoni19,Dobie19,Gomez19,Ackley20,Vieira20,Watson20,Alexander21,deWet21}, which spanned gamma-ray through radio wavelengths and continued for $>250$~days after merger.  No prompt gamma-ray signature was detected despite coverage of the localization region by INTEGRAL, {\it Fermi}, and {\it Swift} \citep{GCN25323,GCN25326,GCN25341}, and observations at X-ray through radio wavelengths did not reveal any likely counterparts.

Here we present a joint analysis of the optical observations of the GW190814 localization region performed by the Gravity Collective, a collaboration consisting of follow-up efforts by the One-Meter Two-Hemisphere (1M2H) collaboration, the Las Cumbres Observatory network, and the Katzman Automatic Imaging Telescope (KAIT) and representing imaging obtained on fourteen 0.7--1~m telescopes across the globe.  We describe our optical searches and follow-up observations of candidates, including optical photometry and spectroscopy, in \autoref{sec:obs}.  In \autoref{sec:candidates}, we discuss our criteria for classifying candidates and conclude that no optical transients discovered from any source are likely counterparts to GW190814.  None of our searches found a viable EM counterpart to GW190814, and so in \autoref{sec:limits} we place limits on the physical nature of any EM counterpart to GW190814.  We compare our limits to models of kilonova, sGRB, or rapidly fading optical emission, from which we determine that we can rule out red kilonovae with ejecta mass and velocity similar to those of AT~2017gfo at $4\times10^{-7}$\% significance and blue kilonovae with ejecta mass $>0.5~M_{\odot}$ (30\% confidence).  We also rule out sGRB afterglows with viewing angles $<17^{\circ}$ \citep[assuming a jet opening angle of 5.2$^{\circ}$ as with GW170817 in][]{Wu18,Wu19} and explosion energies and circumburst densities spanning the range presented by \citet{Fong15}.  We conclude by discussing our overall search and follow-up strategy in \autoref{sec:discussion} and the implications for discovering EM counterparts to GW events in future LVC Observing runs.

All magnitudes presented in this paper are given in the AB system \citep{Oke83}.  Milky Way extinction is derived along the corresponding lines of sight from \citet{Schlafly11}.

\section{Observations}\label{sec:obs}

1M2H coordinated follow-up observations between three 0.7--1~m telescopes in order to search for optical counterparts to GW190814.  The goal of this search and those involving the Las Cumbres Network and KAIT was to localize an optical counterpart to the gravitational-wave event, which was known to resemble an NSBH merger \citep{GCN25333}.  Based on NSBH merger models and their total ejecta masses \citep{Faber06,Ferrari10,Rosswog13}, we inferred that likely counterparts would resemble a kilonova \citep[likely redder than AT~2017gfo; e.g., see][]{Metzger14} or the afterglow from a sGRB.  Based on the localization and luminosity distance to the event provided by the LVC and ultimately representing a volume of $\sim$39,000~Mpc$^{3}$ \citep{Abbott20}, our goal was therefore to localize a counterpart resembling one of these transients within the volume constrained by the event.  

We prioritized 1M2H observations using the open-source code {\tt teglon}\footnote{\url{https://github.com/davecoulter/teglon}}, which examines the LVC localization map and the GLADE galaxy catalog \citep{Dalya18} in order to optimally weight the priority of observing specific parts of the localization region over time.  These observations were prioritized to maximize the likelihood of finding an EM counterpart under the assumption that it occurred in a galaxy in the LVC localization volume.  We also prioritized follow-up observations of viable candidate counterparts with our imaging and spectroscopic resources as described below.

\begin{figure*}
    \centering
    \includegraphics[resolution=300,width=\textwidth]{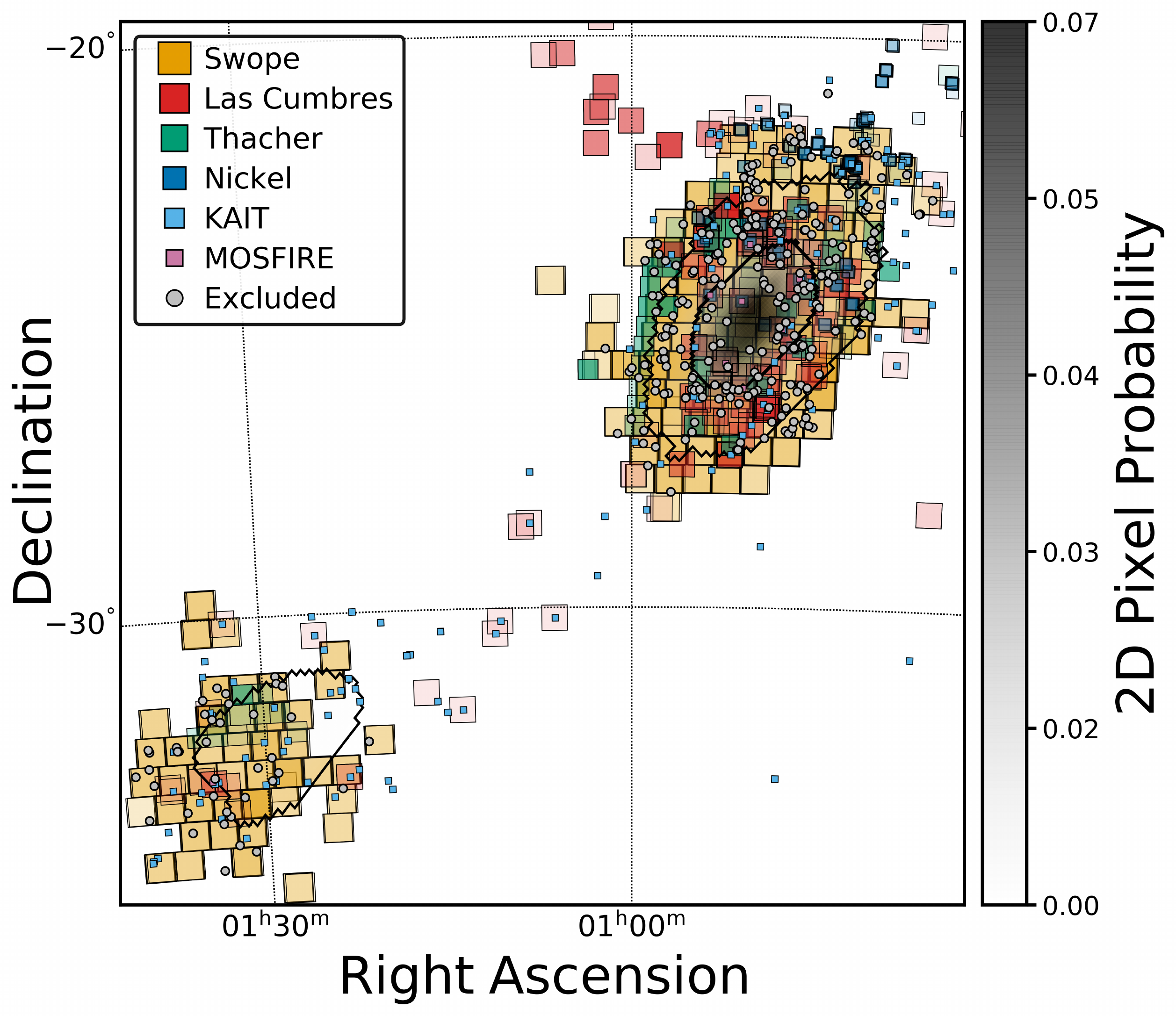}
    \caption{The LVC localization region of GW190814 with our follow-up observations from KAIT (light blue), Las Cumbres (red), Nickel (dark blue), Swope (orange), and Thacher (green) overplotted.  We also show candidate counterparts imaged near the GW190814 localization region and excluded candidates (gray).}
    \label{fig:localization}
\end{figure*}

In addition, we coordinated observations with KAIT and 1~m telescopes in the Las Cumbres Network, all of which targeted galaxies \citep[as in][]{Arcavi17a} in the localization region of GW190814.  Finally, we observed eight galaxies and two counterparts in the localization region of GW190814 with Keck/MOSFIRE, although we were unable to obtain later follow-up imaging using the same instrument and filters.  All observations are described below, and our follow-up images and candidate counterparts are shown relative to the final GW190814 localization region in \autoref{fig:localization}.

\begin{figure*}
\includegraphics[resolution=300,width=\textwidth]{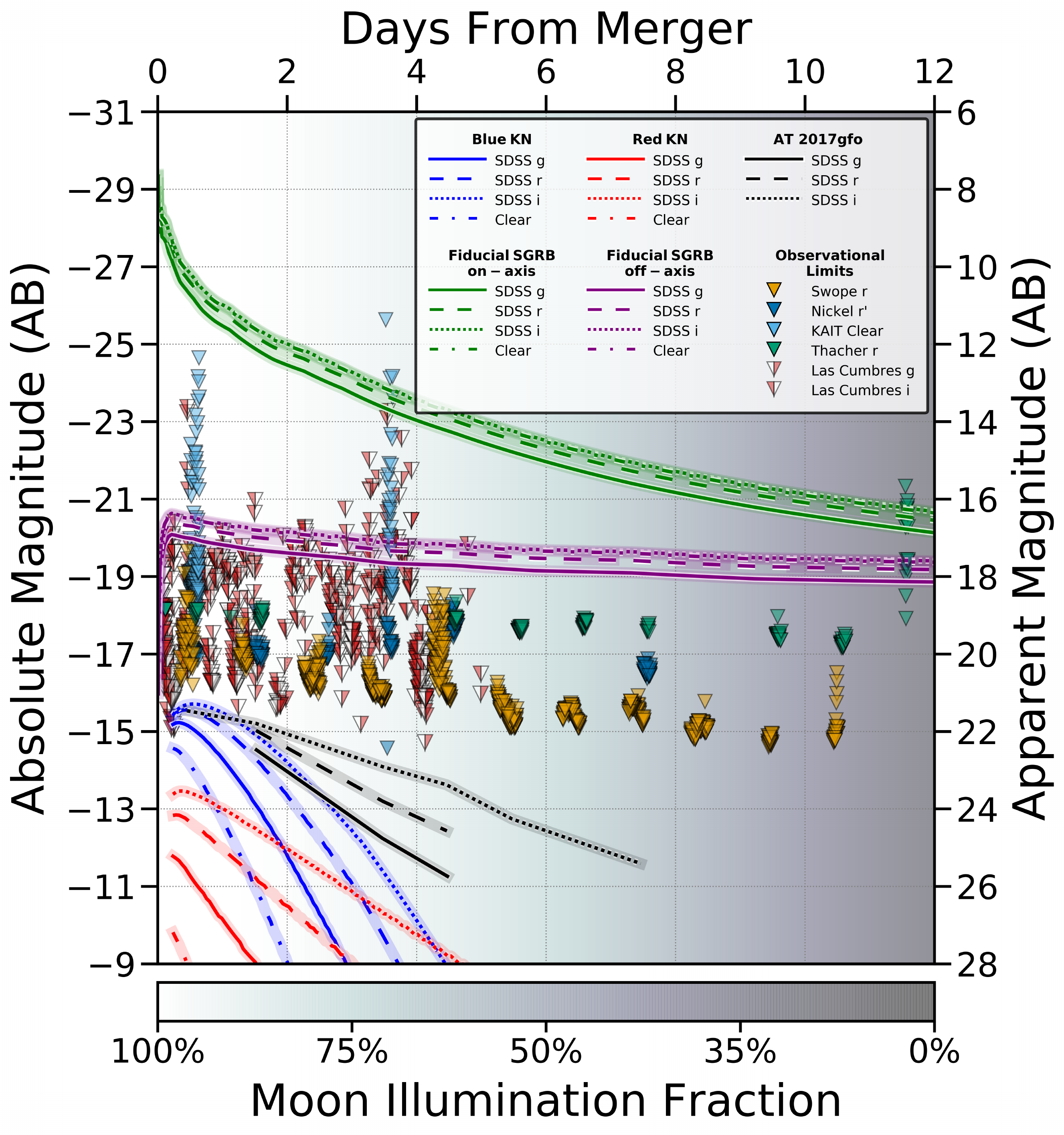}
\caption{The limiting magnitudes from all of our follow-up observations of GW190814 with respect to the time from merger.  We show the absolute magnitudes on the left-hand axis, based on the preferred distance to GW190814 of 241~Mpc, and apparent magnitudes on the right-hand axis.  Each observation is colored by telescope and band as shown in the legend.  We also overplot model light curves for the $gri$ bands based on a blue kilonova with $M_{\rm ej}=0.025~M_{\odot}$, $v_{\rm ej}=0.26$c, $Y_{e}=0.45$, and also a red kilonova with $M_{\rm ej}=0.059~M_{\odot}$, $v_{\rm ej}=0.19$c, $Y_{e}=0.1$, which approximately represent the red and blue components of AT~2017gfo \citep[see \autoref{sec:kilonova} and][]{Drout17,Kilpatrick17}.  Similarly, we show GRB170817A viewed on-axis and off-axis at an angle of $\theta_{\rm obs}=17^{\circ}$ \citep[\autoref{sec:sgrb} and][]{Wu18}.  Finally, we show smoothed $gri$ light curves of AT~2017gfo derived from observations reported by \citet{Drout17}, \citet{Coulter17}, \citet{Arcavi17}, \citet{Cowperthwaite17}, \citet{Troja17}, and \citet{Tanvir17}.}\label{fig:lightcurves}
\end{figure*}

\subsection{Imaging Search and Follow-Up Observations}\label{sec:imaging}

\subsubsection{KAIT}\label{sec:kait}

The 0.76~m Katzman Automatic Imaging Telescope \citep[KAIT;][]{Richmond93,Filippenko01} at Lick Observatory targeted galaxies in the localization region of GW190814 on 15 and 18 Aug.\ 2019, as described by \citet{GCN25353,GCN25437}.  Galaxies were selected from GLADE \citep{Dalya18} according to their $B$-band luminosity, with target priority reweighted by elevation at the time of observation.  All observations were performed in a ``Clear'' filter.  161 galaxies were targeted on 15 Aug.\ 2019 with an additional 52 galaxies on 18 Aug.\ 2019.  All 213 fields were reimaged from 24--25 Aug.\ 2019 to provide templates of the same fields for detailed analysis.

Following standard imaging and photometry procedures \citep[e.g.,][]{Ganeshalingam10,Zheng18}, the images were calibrated and point-spread-function (PSF) photometry was performed using {\tt DAOPHOT} \citep{Stetson87} in {\tt IDL}. The throughput of the KAIT ``Clear'' filter is known to be close to the $R$ band \citep{Li03}, so local AAVSO Photometric All-Sky Survey (APASS) standards \citep{Henden15}\footnote{https://www.aavso.org/apass} were transformed to the Landolt $R$ band \citep{Landolt92} following \citet{Jester05}.  Template images were then subtracted from the 15 and 18 Aug. epochs using a custom {\tt IDL}-based image-subtraction pipeline for PSF convolution.  Finally, the limiting magnitude was estimated in each subtracted image by examining the 3$\sigma$ root-mean-square (RMS) noise within an aperture fixed to the size of the convolved PSF.  These limits are provided in \autoref{tab:observations}.

\subsubsection{Keck/MOSFIRE}\label{sec:mosfireimage}

We targeted 10 fields in a single epoch of target-of-opportunity imaging with the Multi-Object Spectrometer for Infra-Red Exploration \citep[MOSFIRE;][]{mosfire:instr} on the Keck-I 10~m telescope on 15 Aug.\ 2019 as shown in \autoref{tab:observations} and described by \citet{GCN25344}.  Two of these fields targeted candidates within the highest probability regions of GW190814; we observed AT~2019nmd and AT~2019nme \citep[discovered by DESGW;][]{GCN25336,Morgan20}, both of which were undetected and later ruled out as likely minor planets (see \autoref{tab:candidates}). The remaining eight fields targeted galaxies within the inner 50th percentile localization region of GW190814.  The final image mosaics consist of 6.1\arcmin$\times$ 6.1\arcmin~frames centered on the coordinates reported in \autoref{tab:observations}.  All observations consisted of an eight-point dither pattern with 30~s of cumulative exposure time in the $J$ band.

We reduced these data following standard procedures in the MOSFIRE data-reduction pipeline\footnote{\url{https://keck-datareductionpipelines.github.io/MosfireDRP/}} \citep[e.g.,][]{Barro14}.  The images were corrected for dark current and flat-fielded using calibration exposures obtained in the same instrumental configuration.  We then obtained photometry of sources in each image using {\tt DoPhot} and compared these sources to their $J$-band magnitudes in the 2MASS catalog \citep{Cutri03} to calibrate our images.  Comparing to pre-merger 2MASS images of the same fields, we did not detect any transient sources in any of the Keck/MOSFIRE images \citep[as reported by][]{GCN25344}.

The limiting magnitudes reported in \autoref{tab:observations} represent the average RMS sky background inside a single PSF aperture, so they should not be interpreted as the limiting magnitude for any transient sources in our images.  As we were unable to obtain template exposures for these fields, we do not include the Keck/MOSFIRE limits in the analysis of our constraints on EM counterparts to GW190814.

\subsubsection{Las Cumbres}\label{sec:lcogt}

We also observed the localization region of GW190814 with the Las Cumbres Observatory global network \citep{Brown13}, specifically with its 1~m telescopes at the McDonald Observatory in Texas, the Cerro Tololo Interamerican Observatory in Chile, the Siding Spring Observatory in Australia, and the South African Astronomical Observatory. Our pointings were selected based on the galaxy-targeted search and prioritization strategy outlined by \citet{Arcavi17a}. We obtained $300$~s exposures in the $g$ and $i$ bands using the Sinistro cameras mounted on these telescopes, which have a $26' \times 26'$ field of view. Our initial results were reported by \citet{GCN25422}. Image processing was performed by the Las Cumbres Observatory BANZAI pipeline \citep{McCully18} and limiting magnitudes were extracted using {\tt LCOGTSNpipe} \citep{Valenti16}. We used SDSS, PS1, or DECam reference images in the appropriate bands to perform image subtraction using PyZOGY \citep{Zackay16,Guevel17}. The limiting magnitudes were calculated by first estimating the Poisson noise due to the sky using the median absolute deviation of the entire image. Combining the Poisson and read noise, we estimate the 3-$\sigma$ limiting magnitude by inverting the standard signal-to-noise equation. The observations are summarized in \autoref{tab:observations}.

\subsubsection{Nickel}\label{sec:nickel}

We used the Nickel 1~m telescope at Lick Observatory, Mt. Hamilton, California in conjunction with the Direct 2k $\times$ 2k camera ($6.8' \times 6.8'$) to observe galaxies in the localization region of GW190814 on 14--19, 22, 27, and 30 Aug.\ 2019, as well as on 3 and 11 Sep. 2019 (\autoref{tab:observations}).  These images were all obtained in the $r$ band with $180$~s exposures.  Bias-subtraction and flat-fielding were done in {\tt photpipe} \citep{Rest05:photpipe} using calibration frames obtained on the same night and in the same instrumental configuration.  We aligned our images using 2MASS astrometric standards in the image frame and calibrated the images with $r$-band standards obtained from the PS1 DR1 object catalog \citep{Flewelling+16}.  Initial difference imaging was performed using {\tt hotpants} with template images generated from the Dark Energy Camera \citep[primarily DES DR1;][]{DESDR1} and processed using the same pipeline, but our final difference-imaging analysis uses the exposures from Sep. 3 and 11.  In addition, we used a custom version of {\tt DoPhot} \citep{Schechter93} to detect and perform forced photometry on all candidate transient sources.

\subsubsection{Swope}\label{sec:swope}

We observed the localization region of GW190814 with the Swope 1~m telescope at Las Campanas Observatory, Chile from 14 Aug. to 10 Sep. 2019 \citep{GCN25350}.  Each observation is summarized in \autoref{tab:observations}.  The Swope/Direct 4k $\times$ 4k camera on the Swope telescope covers $29.8' \times 29.7'$. We performed all observations in the $r$ band with $120$~s exposures.  Bias-subtraction, flat-fielding, amplifier stitching, image registration, and calibration were all done in {\tt photpipe} following methods described by \citet{Kilpatrick18:16cfr}.  We calibrated each image using $r$-band photometry of stars in the PS1 DR1 object catalog transformed to the Swope natural system using the Supercal method \citep{Scolnic15}.  We performed the difference-imaging procedure described in \autoref{sec:nickel}, and our final analysis uses template exposures obtained from 2--8 Sep. 2019 and listed in \autoref{tab:observations}.

\subsubsection{Thacher}\label{sec:thacher}

The Thacher 0.7~m telescope is a robotic telescope located at the Thacher School Observatory in Ojai, California \citep{Swift18}.  It uses an Andor iKON-L 936 2k$\times$2k imager with a $V$-band optimized back-illuminated chip which translates to a 20.8\arcmin$\times$20.8\arcmin\ field.  We observed the localization region of GW190814 from 15 Aug. to 2 Sep. 2019 with 180~s $r$-band exposures \citep{GCN25351}.  We followed the same calibration and reduction procedure in {\tt photpipe} as described in \autoref{sec:nickel}.  Each frame was calibrated using PS1 DR1 $r$-band standard stars, and we performed initial difference imaging using DECam frames and final analysis with Thacher imaging templates of each field and described in \autoref{tab:observations}.

\subsection{Spectroscopy of Candidates and Hosts}\label{sec:obsspec}

We obtained spectra of candidate EM counterparts to GW190814 and potential host galaxies with Keck, the Shane 3~m telescope at Lick Observatory, and the SOAR 4~m telescope on Cerro Pach\'on, Chile.  Obtaining optical spectra of candidate EM counterparts to GW events can validate whether these sources resemble expectations for kilonovae and sGRB afterglows, as in the case of the initial spectrum of AT~2017gfo obtained $\approx$12~hr after the GW event \citep{GCN21547,Shappee17}.  Similarly, optical spectra of transients unlikely to be associated with a GW event, such as supernovae, can be used to rule out a candidate counterpart.  Finally, spectra provide a redshift to the event, enabling us to rule out candidate counterparts based on a volumetric cut in the context of the LVC localization region and luminosity distance. All of our spectroscopic observations are summarized in \autoref{tab:spectra}.  Below we detail our observation and reduction procedure for each telescope.

\subsection{Keck/DEIMOS}\label{sec:deimos}

We obtained a spectrum of the GW190814 counterpart candidate AT~2019osy on 28 Aug. 2019 using the DEep Imaging Multi-Object Spectrograph (DEIMOS) \citep{2003SPIE.4841.1657F} on the Keck-II 10 m telescope. Two 900 second exposures were taken with a 1\arcsec long slit, 600ZD grating, and the GG455 long-pass order blocking filter.

The spectra were reduced with the {\tt PypeIt} version 1.3.3 \citep{pypeit:joss_arXiv, pypeit:zenodo} with the standard reduction.  {\tt PypeIt} was also used to flux calibrated the spectra with a standard star observed on the same night and to coadded the observations into a single spectrum.

\subsubsection{Keck/MOSFIRE}\label{sec:mosfirespec}

\begin{figure}
    \centering
    \includegraphics[width=0.49\textwidth]{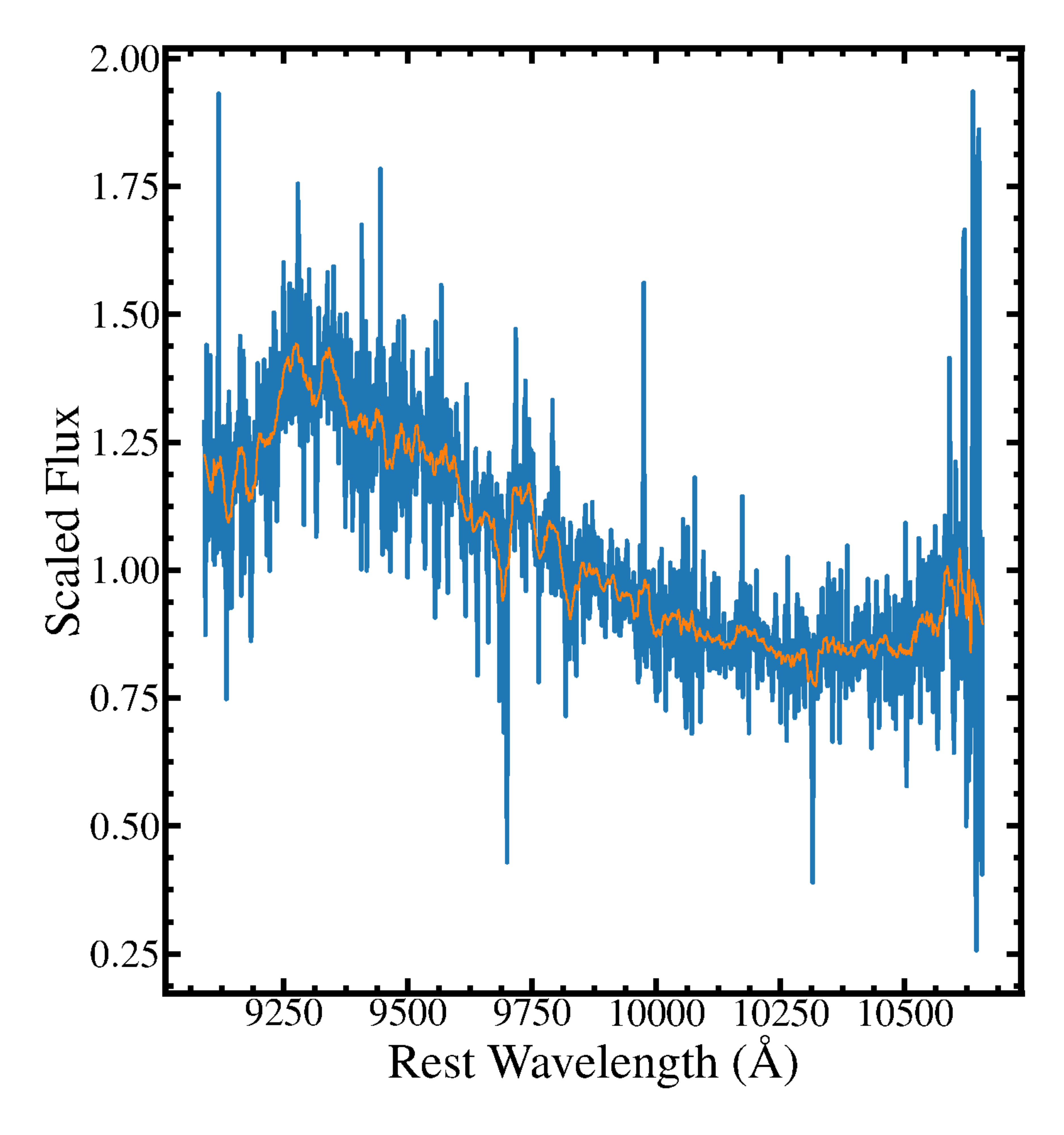}
    \caption{Keck/MOSIFRE spectrum of the GW190814 candidate counterpart AT~2019nra ($z=0.269\pm0.029$) obtained on 18 Aug.\ 2019 as reported by \citet{GCN25395}.  The spectrum is largely featureless apart from a single intermediate-width feature at a rest wavelength of 9300~\AA, although this wavelength does not correspond to any known features.}
    \label{fig:mosfire}
\end{figure}

We used Keck/MOSFIRE to observe the candidate counterpart to GW190814 AT~2019nrm on 18 Aug.\ 2019 as reported by \citet{GCN25395}. We obtained $4\times120$~s (ABBA nod pattern) $J$-band ($\sim$11,500--13,500~\AA) spectroscopic observations, using the 0.7\arcsec\ slit. The spectra were reduced with the MOSPY Data Reduction Pipeline\footnote{https://keck-datareductionpipelines.github.io/MosfireDRP/} and calibrated with standard-star observations, taken on the same night and in the same instrumental configuration.  The final spectrum is shown in \autoref{fig:mosfire}.

\subsubsection{Lick Shane/Kast}\label{sec:kast}

We observed candidate EM counterparts and host galaxies with the Kast double spectrograph \citep{Miller93} on the Shane 3~m telescope on 26 and 31 Aug., 2, 5, and 21 Sep., and 8 Oct. 2019 as shown in \autoref{tab:spectra}.  All observations were taken with the 452/3306 grism (blue arm) and the 300/7500 grating (red arm), using the 2\arcsec\ wide slit, covering approximately 3200--10,500~\AA\ in the combined blue-side and red-side spectra. The spectra were reduced with standard {\tt IRAF}\footnote{{\tt IRAF} is distributed by the National Optical Astronomy Observatory, which is operated by the Association of Universities for Research in Astronomy (AURA), Inc., under a cooperative agreement with the National Science Foundation.} CCD-processing and spectrum-extraction procedures, and our own {\tt IDL} routines for flux calibration and telluric-line removal, using the well-exposed continua of spectrophotometric standard stars. Details on the spectroscopic reduction procedure are outlined by \citet{Silverman2012MNRAS}.

\subsubsection{SOAR/Goodman}\label{sec:goodman}

We used the Goodman spectrograph on the Southern Astrophysical Research (SOAR) 4~m telescope to observe candidate EM counterparts and host galaxies of GW190814 on 17 Aug.\ and 1 Sep.\ 2019.  All observations were performed using the 400~line mm$^{-1}$ M1 (3000--7000~\AA) grating in conjunction with the 1.07\arcsec\ wide slit.  We reduced all spectra following the same procedures as for the Shane/Kast data.  These spectra are shown in \autoref{fig:spectra} and discussed in \autoref{sec:candidates}.

\begin{deluxetable*}
{lccccc}
\tabletypesize{\scriptsize}
\tablecaption{Spectra of Candidates and Host Galaxies\label{tab:spectra}}
\tablewidth{5pt}
\tablehead{
\colhead{Name} & 
\colhead{Observation Date} & 
\colhead{Source} & 
\colhead{Type} & 
\colhead{$z$} & 
\colhead{Ref.}}
\startdata
2019noq & 2019-08-21 & Goodman & II & 0.0591$\pm$0.0003 & 1,4 \\
2019npf & 2019-10-03 & Kast    & -- & 0.1630$\pm$0.0001 & 2 \\
2019nph & 2019-10-31 & Kast    & -- & 0.2689$\pm$0.0003\tablenotemark{\footnotesize a} & 2 \\ % RVSAO=m31_a_temp.fits 4.38 80629.9 77.497 bad
2019npv\tablenotemark{\footnotesize b} & 2019-08-26 & Magellan & Ibc &  0.0560$\pm$0.0001 & 1,3,4,11 \\
2019npw & 2019-08-27 & Goodman & II & 0.1494$\pm$0.0001 & 3,5,6 \\
2019nqc\tablenotemark{\footnotesize b} & 2019-08-23 & SALT & II & 0.0780$\pm$0.0001 & 1,4,5,7,11 \\
2019nqg & 2019-08-31 & Kast    & -- & 0.1706$\pm$0.0001 & 5 \\
2019nqq & 2019-08-17 & Goodman & II & 0.0711$\pm$0.0001 & 1,4,8,11 \\
2019nqr & 2019-08-17 & Goodman, FLOYDS & II & 0.0832$\pm$0.0001 & 1,4,9,10,11 \\
2019nqx & 2019-10-08 & Kast    & -- & 0.2792$\pm$0.0003\tablenotemark{\footnotesize a} & 12 \\ % RVSAO=m31_k_temp.fits 3.47 83697.3 91.454 pretty bad
2019nrd & 2019-08-31 & Kast    & -- & 0.2472$\pm$0.0003 & 12 \\
2019nra & 2019-09-21 & Kast    & -- & 0.2971$\pm$0.0002\tablenotemark{\footnotesize a} & 12,13 \\ % RVSAO=m31_k_temp.fits 2.89 89069.1 67.827 bad, but reasonable H-alpha guess
2019nra & 2019-08-18 & MOSFIRE & ?  &   & 12,13 \\
2019nrb & 2019-10-09 & Kast    & -- & 1.7953$\pm$0.0011\tablenotemark{\footnotesize a} & 12 \\ % RVSAO=hemtemp0.0.fits 4.85 538243.7 29.813 bad
2019nrp & 2019-11-06 & Kast    & -- & 0.0107$\pm$0.0002\tablenotemark{\footnotesize a} & 12 \\ % RVSAO=sptemp.fits 3.54 3222.71 70.155 bad, might be Halpha there but I don't see template matches
2019nrv & 2019-10-09 & Kast    & -- & 0.0531$\pm$0.0003\tablenotemark{\footnotesize a} & 12 \\ % RVSAO=habtemp90.fits 2.88 15916.7 90.832 bad
2019nte & 2019-08-22 & Goodman & -- & 0.0706$\pm$0.0001 & 4,11,14,15 \\
2019nte & 2019-10-08 & Kast    & -- & 0.0700$\pm$0.0004\tablenotemark{\footnotesize a} & 4,11,14,15 \\ % RVSAO=eatemp 20989.147 111.182 1.38 bad, looks like broad line AGN
2019ntn & 2019-08-21 & Goodman & Ia-CSM & 0.1001$\pm$0.0002 & 1,4,16 \\
2019ntp & 2019-09-01 & Goodman & Ia & 0.1141$\pm$0.0001 & 1,4,15,16,17 \\
2019ntr & 2019-08-29 & Goodman & II & 0.2185$\pm$0.0001 & 1,4,11,15,16,18\\
2019nts & 2019-08-31 & Kast    & -- & 0.1931$\pm$0.0003\tablenotemark{\footnotesize a} & 1,16 \\ % RVSAO=m31_f_temp.fits 3.37 57896.3 78.436 ???
2019nul & 2019-08-31 & Kast    & -- & 0.0985$\pm$0.0002 & 1,4,16,19 \\
2019num & 2019-08-27 & Goodman & IIb& 0.1274$\pm$0.0001 & 1,4,15,16 \\
2019nuo & 2019-10-03 & Kast    & -- & 0.1151$\pm$0.0001 & 15,19 \\
2019nun & 2019-09-05 & Kast    & -- & 0.1319$\pm$0.0002 & 1,16,19 \\
2019nur & 2019-08-26 & Kast    & -- & 0.1394$\pm$0.0002 & 16 \\
2019nuu & 2019-08-26 & Kast    & -- & 0.2106$\pm$0.0002 & 16 \\
2019nwt & 2019-08-31 & Kast    & -- & 0.2458$\pm$0.0003 & 20 \\
2019nyv & 2019-09-05 & Kast    & -- & 0.0410$\pm$0.0004\tablenotemark{\footnotesize a} & 1,21 \\ %RVSAO=eatemp.fits 3.07 12278.0 124.254 bad
2019nyz & 2019-08-26 & Kast    & -- & 0.4146$\pm$0.0002 & 21 \\
2019nzg & 2019-08-26 & Kast    & -- & 0.2132$\pm$0.0002 & 21 \\
2019nzm & 2019-09-21 & Kast    & -- & 0.2143$\pm$0.0002 & 21 \\
2019nzn & 2019-09-03 & Kast    & -- & 0.1716$\pm$0.0001 & 19,21 \\
2019nzr & 2019-09-02 & Kast    & -- & 0.2549$\pm$0.0002 & 1,11,15,22 \\
2019nzs & 2019-10-03 & Kast    & -- & 0.2261$\pm$0.0002 & 22 \\
2019oat & 2019-09-21 & Kast    & -- & 0.1978$\pm$0.0003\tablenotemark{\footnotesize a} & 21 \\ %RVSAO=m31_f_temp.fits 2.63 59322.2 87.887 bad
2019oaz & 2019-09-03 & Kast    & -- & 0.1987$\pm$0.0003 & 21 \\
2019obb & 2019-09-05 & Kast    & -- & 0.3157$\pm$0.0002 & 22 \\
2019obc\tablenotemark{\footnotesize b} & 2019-08-23 & GTC & Ia & 0.216$\pm$0.005   & 1,4,11,15,22,23 \\
2019odc & 2019-08-26 & Kast    & -- & 0.0551$\pm$0.0002 & 1,4,11,15,24 \\
2019ofb & 2019-08-31 & Kast    & -- & 0.1185$\pm$0.0001 & 1,25 \\
2019omx & 2019-08-29 & Goodman & -- & 0.1645$\pm$0.0001 & 1,4,11,15,18,26 \\
2019onj & 2019-09-03 & Kast    & -- & 0.0665$\pm$0.0001 & 1,11,15,27,28 \\
2019osy & 2019-08-28 & DEIMOS  & -- & 0.0738$\pm$0.0003 & 1,29,30,31 \\  
\enddata
\tablecomments{Our Keck/DEIMOS, Keck/MOSFIRE, Shane/Kast, and SOAR/Goodman spectra are described in \autoref{sec:obsspec}.  All dates are UTC. The spectral types for transient spectra are noted along with spectroscopic redshifts.  Where we have only measured the host-galaxy redshift, we note the spectral type of the spectrum as ``--.'' We provide references relevant to discovery and classification of each candidate as follows: (1) \citet{Ackley20} , (2) \citet{2019TNSTR1496}, (3) \citet{2019TNSCR1643}, (4) \citet{Andreoni19}, (5) \citet{2019TNSTR1507}, (6) \citet{GCN25484}, (7) \citet{2019TNSCR1652}, (8) \citet{2019TNSTR1508}, (9) \citet{2019TNSTR1509}, (10) \citet{2019TNSCR1522}, (11) \citet{Morgan20}, (12) \citet{2019TNSTR1516}, (13) \citet{GCN25395}, (14) \citet{2019TNSTR1526}, (15) \citet{GCN25486}, (16) \citet{2019TNSTR1525}, (17) \citet{GCN25596}, (18) \citet{GCN25540}, (19) \citet{GCN25445}, (20) \citet{2019TNSTR1536}, (21) \citet{2019TNSTR1562}, (22) \citet{2019TNSTR1563}. (23) \citet{GCN25543}, (24) \citet{2019TNSTR1579}, (25) \citet{2019TNSTR1584}, (26) \citet{2019TNSTR1604}, (27) \citet{2019TNSTR1615}, (28) \citet{GCN25526}, (29) \citet{GCN25822}, (30) \citet{GCN25801}, (31) \citet{GCN25487}.}
\tablenotetext{a}{Spectra did not meet our cross-correlation height-to-noise ratio ($r$) threshold $r>4$.  See discussion in \autoref{sec:specz}.}
\tablenotetext{b}{Spectra of these objects are not presented in this publication, but their classifications and redshifts are used in our analysis and can be found in the references given.}
\end{deluxetable*}

\subsection{Identification of Transients}\label{sec:transients}

Follow-up imaging from 1M2H, KAIT, and Las Cumbres was used to identify transients that we consider candidate counterparts to GW190814 after we performed difference imaging.  Here we summarize the methods used for elevating transient sources to candidate counterparts within each set of imaging.  All photometry of transient sources used in this analysis, including photometry from outside sources \citep[e.g.,][]{Andreoni19,Ackley20,Morgan20}, is summarized in \autoref{tab:cand-photometry}.

\subsubsection{1M2H}

As described above, all Nickel, Swope, and Thacher difference imaging was analyzed through the {\tt photpipe} difference imaging and analysis pipeline.  After difference imaging through {\tt hotpants} and identification of transient sources using a custom version of {\tt dophot}, {\tt photpipe} cuts transient sources based on pixel-level statistics including the relative fraction of positive, negative, and masked (i.e., saturated) pixels within the PSF aperture and the extendedness of the source PSF relative to the PSF passed to {\tt dophot}.  In general, these criteria are relaxed in order to avoid cutting a significant fraction of the transient source catalog for each image, but these cuts naturally result in a loss in detection efficiency for each image (typically $\approx$3\% of sources on average).  We perform the same cuts during fake star injection, and so we account for this detection efficiency in our limiting magnitude for each image.  On average, we detect 2 sources in each Nickel image, 12 sources in each Swope image, and 10 sources in each Thacher image.

After we construct a transient catalog for each image, sources are crossmatched across all catalogs using a search radius of 2~pixels, that is, transient sources within 2~pixels of the average coordinate of a previously identified ``cluster'' of detections are considered to be the same source.  Sources are elevated as candidates if they are 1) detected in at least two images with $S/N>3$, or 2) in a single image with $S/N>10$.  We obtain final photometry for every cluster of sources by taking the signal-to-noise weighted average position of each cluster and running forced photometry on this position with {\tt dophot}.  Finally, we visually inspect all candidate transients to validate that they are not due to poor image quality or other non-astrophysical contaminants.  All 1M2H candidates are crossmatched against known minor planets before we consider them as viable candidate counterparts to GW190814.

\subsubsection{KAIT}

KAIT images were analyzed through part of our custom-developed {\tt LOSSPhotPypeline}\footnote{https://github.com/benstahl92/LOSSPhotPypeline} \citep{Stahl19}, which adopts the {\tt ISIS} package\footnote{http://www2.iap.fr/users/alard/package.html} \citep{Alard98} for image subtraction. After difference imaging, identification of transient sources was based on several parameters (e.g., PSF, FWHM, mag, S/N) extracted from the original image, the template image, and the residual image using {\tt SExtractor}\footnote{https://github.com/astromatic/sextractor} \citep{sextractor}.

Candidates passing the customized criteria were logged on a web-based list, and then further visually checked by multiple people to eliminate any non-astrophysical contaminants. Similar to 1M2H, all  of our candidates were crossmatched against known minor planets before we considered them as viable candidate counterparts. If no valid candidate was found, the limiting magnitude was estimated by examining the 3$\sigma$ RMS noise averaged through several locations across the entire image.

\section{Classification of Candidates}\label{sec:candidates}

Although no optical candidate counterparts to GW190814 were definitively identified as being similar to kilonovae or sGRBs in the extensive follow-up observations and analysis of the event \citep[e.g., in the optical or radio;][]{Andreoni19,Dobie19,Gomez19,Thakur20,Morgan20}, limits on the physical nature to any counterpart rely on the assumption that all transient sources identified in the aftermath of GW190814 have been ruled out as being associated with the GW detection.  In the following analysis, we discuss all known candidate optical transients that we consider potentially associated with GW190814.  We assess the likelihood that one of these transient sources was the EM counterpart to the merger using the localization region and time of merger, candidate spectroscopy, premerger variability, host-galaxy associations, and candidate photometry.  The following criteria describe our classification procedure and rationale.

\begin{enumerate}

    \item {\bf Localization}: We consider publicly-reported candidates and those discovered by our own surveys that are within the 99th percentile credible region provided by the LVC, a total area of 48.7~deg$^{2}$ and \Ncandidate\ viable candidates (\autoref{fig:localization}).  This can theoretically result in missing the actual counterpart, but we expect it to occur in only 1\% of cases.  Furthermore, we find that for the inner 95th percentile credible region (26.8~deg$^{2}$) there are 214 viable candidates while for the 99.5th percentile (58.2~deg$^{2}$) there are 290 candidates.  The fact that the number of candidates does not scale as the total search area reflects the shallower search depth of the lower-probability areas in the GW190814 localization region.  Therefore, we are confident that we do not rule out any known, high-probability transients on the basis of localization.

    \item {\bf Time from Merger}: We restrict our analysis to candidates discovered within 14~days after the LVC detected the GW190814 merger signal. This will only result in false negatives if the optical counterpart to GW190814 has a rise time significantly longer than known kilonovae and sGRBs such as AT~2017gfo \citep[implying, for example, a kilonova with an extremely large ejecta mass and optical opacity or a highly off-axis GRB;][]{Rossi02,Yamazaki03,Ryan15,Kasen14,Metzger16,Kasen17}.  The lack of a prompt gamma-ray counterpart implies that if there was a GRB170817A-like GRB \citep{Troja17,Mooley18}, it was likely off-axis and thus not detectable by INTEGRAL, {\it Fermi}, or {\it Swift} \citep[despite coverage of the region as described by][]{GCN25323,GCN25326,GCN25341}.  However, while the rise time for a GRB afterglow scales relative to the viewing angle $\theta_{\rm obs}$ and jet opening angle $\theta_{0}$ as $\theta_{\rm obs}/\theta_{0}$, optical follow-up observations would likely be insensitive to a counterpart observed $>$14~days after merger \citep[see models by][]{Lazzati17,Murguia-Berthier17}.  A delayed rise in optical luminosity could also occur if the merger occurred in an evacuated circumburst medium \citep[proposed by][for GW170817]{ramirezruiz19}, similar to the Galactic pulsar J1913+1102 \citep{Lazarus16}. However, for plausibly detectable EM counterpart models, the rise time is significantly shorter than 14~days, and so we are confident that this restriction is conservative.
    Furthermore, we find that if we restrict our analysis to candidates discovered only within the first 10~days post-merger while keeping our localization cut the same, there are 270 viable candidates, whereas if we cut at 21~days there are 293.  Similar to localization, this reflects the fact that follow-up observations decreased significantly in depth and cadence at $>$10~days post-trigger.  Criteria \#1 and \#2 define our initial sample of \Ncandidate\ candidate counterparts, including \Ninternal\ candidates identified in our own follow-up program (\autoref{tab:candidates}).

    \item {\bf Coincidence with minor planets}: Candidates that are coincident ($<20$\arcsec) with minor planets at the time of observation and as reported by the Minor Planet Center\footnote{\url{https://minorplanetcenter.net}}, and are not detected in multiple epochs of imaging separated by $>30$~min, are ruled out.  For candidates discovered by 1M2H, KAIT, and Las Cumbres, we perform this check before considering a transient as a viable EM counterpart to GW190814, and so minor planets are not reported in \autoref{tab:candidates}.  For all other publicly-reported transients, this check is typically performed before a candidate is reported, but some reported candidates were reclassified as minor planets, for example, in \citet{Andreoni19} and \citet{Morgan20}.  Overall, we rule out \Nmpc\ objects based on coincidence with minor planets.

    \item {\bf Coincidence with known stars}: We checked each candidate for coincidence ($<1$\arcsec) with nearby stars in the {\it Gaia} DR2 catalog \citep[i.e., those that have parallax or proper motion measured at $>3\sigma$;][]{GaiaDR2}.  Transient detections can potentially arise from stellar variability, and so are not expected to be associated with GW190814.  However, \Ngaia\ objects are ruled out based solely on coincidence with {\it Gaia} DR2 stars.

    \item {\bf Spectroscopic classification}: Candidates with spectra that resemble known classes of transients \citep[i.e., those with template spectra in SNID;][]{Blondin07} are ruled out as counterparts to GW190814.  To perform this analysis, we consider spectra presented in publications on GW190814 \citep[e.g.,][and references therein]{Dobie19,Andreoni19b,Gomez19,Ackley20,Watson20}.  We only consider spectra that we can definitively classify as supernovae or other well-known classes of transients that are not thought to be associated with NS mergers.  For example, spectra that resemble ``blue continuum'' or galaxy emission are not ruled out, as AT~2017gfo was initially very blue \citep{Drout17,Shappee17,Kilpatrick17} and spectra dominated by galaxy emission imply non-detection of the transient.  Thus, the likelihood of false negatives is negligible.  We rule out \Nclass\ candidates based solely on spectroscopic classification, which includes all sources with classifications in \autoref{tab:spectra}.

    \item {\bf Pre-merger variability}: Candidates with detections in transient surveys from before the merger are ruled out, which we derive from crossmatching ($<1$\arcsec) to the Pan-STARRS DR2 Detection Catalog \citep[see the description by][]{Flewelling+16}, an available light curve in the ASAS-SN Photometry Database\footnote{\url{https://asas-sn.osu.edu/}} \citep{Shappee14b,Jayasinghe19}, or the Asteroid Terrestrial-impact Last Alert System forced photometry server \citep{ATLAS-forced}.  Models for merging NSBH systems do not typically predict any significant premerger optical emission, for example due to accretion from a circumbinary disk \citep{Joss84,2018ApJ...862L...3S}; however, we further consider active galactic nucleus (AGN) variability \citep[as in][]{Graham20} as a possible counterpart to GW190814 in \autoref{sec:agn}.  We define variability as multiple premerger detections in a single band with a significant ($>$3 $\sigma$) change in brightness.  We rule out \Nvariable\ objects based solely on premerger variability, each of which is discussed below.  We consider candidates ruled out by this criterion or any of the preceding steps to be ``strongly ruled out'' (as described in \autoref{tab:candidates}), comprising \Nstrong/\Ncandidate\ (\Fstrong) of our sample.

    \item {\bf Spectroscopic redshifts}: The LVC constrain the distance to GW events using signal amplitude \citep{Abadie11}.  For GW190814, this resulted in a luminosity distance of $241\substack{+41\\-45}$~Mpc ($z=0.056\pm0.010$ assuming H$_{0}=70~\text{km s}^{-1}~\text{Mpc}^{-1}$; \citealt{Abbott20}).  Candidates found in host galaxies that are outside the 99th percentile credible volume (roughly $D_{\mathrm{mean}} \pm 2.58\,D_{\mathrm{std}}$) as determined from spectroscopic redshifts are ruled out.  This criterion can result in false negatives when there is a chance coincidence between a candidate and assumed host galaxy.  However, the mean number density of galaxies with luminosity $>L_{\star}$ in the local Universe is approximately $6\times10^{-3}~\text{Mpc}^{-3}$ \citep{Schechter76,Bell03}, implying roughly one galaxy per 85~arcmin$^{2}$.  Where host identification is made, the median candidate-host separation is $\sim$2\arcsec, and so the chance coincidence with a random host galaxy in the GW190814 volume is 0.004\% (1 per 25,000 candidates), or a $<$1\% chance of occurring at most once for all \Ncandidate\ of our candidates.  To further reduce the likelihood of chance coincidence, we require that for a galaxy with a spectroscopic or photometric redshift, the projected separation between the host galaxy and transient is $<$300~kpc \citep[assuming {\it Planck} 2016 cosmology;][]{Planck15}.  We derive spectroscopic redshifts from public databases, our own observations as described below, and observations in the literature.  Therefore, we do not think that false negatives from candidates ruled out this way are significant.  We rule out \Nspecz\ candidates based solely on spectroscopic redshifts outside the 99\% credible volume.

    \item {\bf Photometric redshifts}: We repeat the previous step using photometric redshifts from the 2MASS Photometric Redshift \citep[2MPZ;][]{Bilicki13}, Pan-STARRS1 Source Types and Redshifts with Machine learning (PS1-STRM) catalogs \citep{PS1redshift}, and the Photometric Redshifts for the Legacy Surveys \citep{Zhou20}.  We choose the ``best'' host-galaxy association and photometric redshift by crossmatching galaxies between each catalog and picking the host galaxy with the smallest projected separation (in kpc), and then choosing the photometric redshift with the smallest relative uncertainty $\sigma_{z}/z$.  All of these catalogs are trained using machine-learning approaches, and on 2MASS, {\it WISE}, and SuperCOSMOS photometric data for the former and PS1 data for the latter.  These result in median redshifts $z<0.1$ for both catalogs, making them ideal for low-redshift analysis of transients in the LVC volume.  We rule out \Nphotoz\ candidates based on photometric redshift, representing \fphotoz\ candidates overall.

    \item {\bf Photometric evolution}: It is now known that NS mergers result in kilonovae and sGRBs \citep[][]{Abbott17:grb,Kasen17}, the optical properties of which have been constrained by sGRB follow-up observations at higher redshifts and the discovery of AT~2017gfo \citep{Arcavi17,Drout17,Kilpatrick17}.  Apart from the most extreme physical scenarios and geometries, theoretical models and observations imply that EM counterparts to NS mergers are quite faint, fade rapidly, and typically have colors $g-r>-1.0$~mag (all ``red'' kilonova models considered by \citealt{Kilpatrick17} peak at $>-18$~mag, fade more rapidly than $0.1$~mag~day$^{-1}$ in optical bands, and have $g-r > 1.0$~mag at all epochs).  Therefore, we can rule out candidates that would be luminous at the distance to GW190814, have rising or slowly fading light curves, or are bluer than expected for sGRB afterglows.  By using these limits and considering candidates with arbitrarily faint, rapidly evolving, or red light curves, we avoid false negatives due to host extinction.  We rule out \Nphot\ candidates by photometric cuts, which represents all candidates not ruled out by previous cuts.  Candidates ruled out only by these criteria are ``weakly ruled out.''

\end{enumerate}

To systematically assess the viability of each candidate counterpart, we determine the extent to which all candidates match each of these criteria in \autoref{tab:candidates}.  We start from our base sample of \Ncandidate\ candidates derived from publicly reported transients and candidates discovered in our follow-up imaging, all of which satisfy the localization criterion and were discovered in imaging taken within 2~weeks of the GW190814 merger.  We conclude that there are no viable candidate counterparts, although 28 (14\%) of all candidate counterparts are ``weakly ruled out'' by the photometric evolution criteria.  In the remainder of this section, we discuss individual candidates and whether they are likely counterparts to GW190814 in the context of our overall classification scheme.  We summarize the steps we use to rule out candidates in \autoref{tab:cand-analysis}, including the number of candidates that can be ruled out at each step ignoring all previous steps.

\begin{deluxetable*}
{llccc}
\tabletypesize{\scriptsize}
\tablecaption{Analysis of Public and Internal Candidate Counterparts to GW190814\label{tab:cand-analysis}}
\tablewidth{0pt}
\tablehead{
\colhead{Step} &
\colhead{Criterion} &
\colhead{Candidates} &
\colhead{Candidates Cut} &
\colhead{All Candidates Flagged\tablenotemark{\footnotesize a}}}
\startdata
0 & Within Spatial (99th percentile area) and Temporal Cut ($<$2 weeks from merger)       & \Ncandidate & --         & --      \\
1 & Not $<$20\arcsec\ from minor planet and not in multiple epochs separated by $>$30~min & 178         & \Nmpc      & \Nmpc   \\
2 & Not $<$1\arcsec\ from star with parallax in Gaia DR2                                  & 178         & \Ngaia     & 0       \\
3 & Not spectroscopically classified as an unassociated transient                         & 166         & \Nclass    & \Nclass \\
4 & No evidence for pre-merger variability                                                & 164         & \Nvariable & 2       \\
\hline
5 & Spectroscopic redshift not outside the 99th percentile credible volume                & 133         & \Nspecz    & 39      \\
6 & Photometric redshift not outside the 99th percentile credible volume                  & 28          & \Nphotoz   & 127     \\
\hline
7 & Not photometrically classified as an unassociated transient                           & 0           & \Nphot     & 126     \\
\enddata
\tablenotetext{a}{This column indicates the number of candidates that would be cut at this stage if we disregard all previous stages.}
\end{deluxetable*}

\subsection{Coincidence with Minor Planets}\label{sec:mpc}

The primary interloping transients in our own follow-up imaging are minor planets, which result from asteroids with proper motion that is high enough to appear in only a single epoch of imaging but low enough not to appear intrinsically extended in a 2--5~min exposure or show up in multiple exposures separated by $<$10~min.  \citet{Andreoni19} attempt to rule out these transients by taking multiple, nonconsecutive exposures when possible, but several objects in the Minor Planet Center database were reported when transients were detected only in a single epoch or the follow-up image occurred soon after discovery.  In particular, we note the following transients in the Transient Name Sever that were reported with 2 detections but a short period of time between the two epochs (given here in parentheses next to each candidate): AT~2019nmd (1.4~min), AT~2019nme (1.4~min), AT~2019nqh (single epoch), AT~2019nri (8.6~min), AT~2019nrq (2.9~min), AT~2019nrt (2.9~min), AT~2019nsd (2.9~min), AT~2019nsl (2.9~min), and AT~2019nsn (2.9~min) \citep{Andreoni19b,Morgan20}.  In both \citet{Andreoni19b} and \citet{Morgan20}, these candidates were subsequently revised to indicate that they are in fact minor planets as we note here.

Given the coincidence ($<$20\arcsec) between these transients and minor planets reported in \autoref{tab:candidates} as well as the correlation with transients detected in a single epoch or at most two epochs separated by $<$30~min, we rule out all of these sources as likely minor planets and thus unassociated with GW190814.  As described in \autoref{sec:transients}, this cut was performed a priori for our own data, and so no internal candidates from 1M2H, KAIT, or Las Cumbres are considered likely minor planets.

\subsection{Spectroscopy of Candidate Counterparts}\label{sec:candspec}

\begin{figure*}
    \centering
    \includegraphics[width=0.95\textwidth]{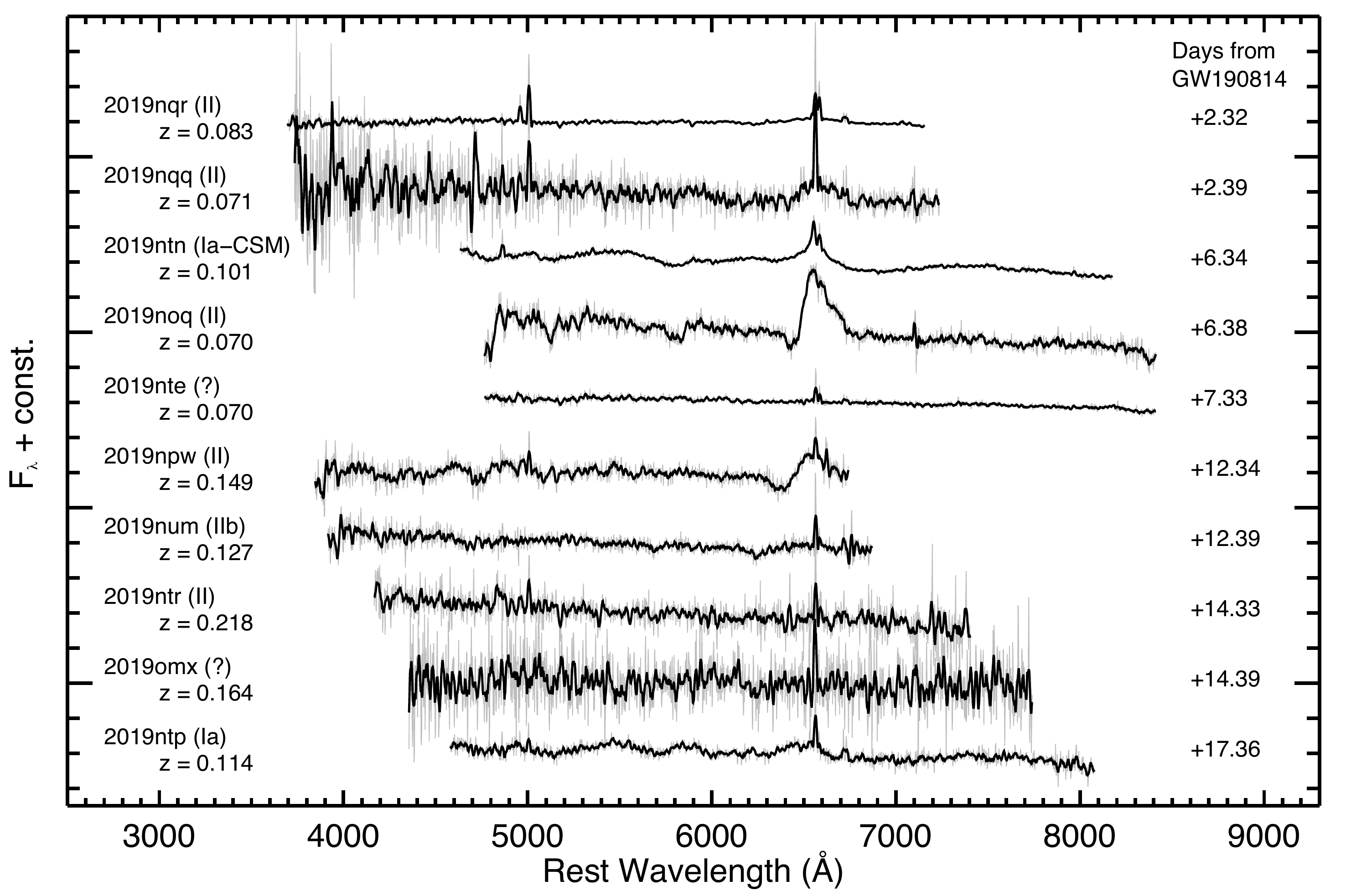}
    \caption{Spectra of candidate optical counterparts to GW190814 obtained with SOAR/Goodman.  No spectra are classified as likely kilonovae or sGRBs, and thus we determine based on spectroscopic classification that most objects in this sample are unassociated with GW190814.}
    \label{fig:spectra}
\end{figure*}

Our Keck, Shane, and SOAR spectra of GW190814 candidate counterparts were obtained from 17 Aug.\ to 1 Sep.\ 2019 and definitively rule out some counterparts as being likely kilonovae or sGRB afterglows.  Below we examine spectroscopic classifications of individual candidates inferred from spectra discussed in \autoref{sec:obs} and in other publications.  Our classifications are performed using SNID \citep{Blondin07}, from which we derive redshift and spectral type.  All of our candidate spectra are shown in \autoref{fig:spectra}.

{\it AT~2019noq}: We obtained a SOAR/Goodman spectrum of the candidate AT~2019noq \citep[also PS19epf, discovered on 15 Aug.\ 2019 by][]{GCN25356} on 21 Aug.\ 2019 as reported by \citet{GCN25423}.  The spectrum is consistent with a Type~II SN at $z=0.07$, and so we do not consider AT~2019noq to be a likely counterpart to GW190814.

{\it AT~2019ntn}: AT~2019ntn was discovered on 18 Aug.\ 2019 as part of DECam-GROWTH follow-up observations \citep{GCN25393,GCN25394,Andreoni19}.  We obtained a spectrum with SOAR/Goodman on 21 Aug.\ 2019.  This spectrum was reported as similar to that of a Type Ia-CSM or Ia/IIn at $z=0.1$ by \citet{GCN25423}.  We find similar results, and so we rule out any association between AT~2019ntn and GW190814.

{\it AT~2019npw}: AT~2019npw was initially reported by \citet{GCN25362} at $i=20.5$~mag and characterized as offset from its likely host galaxy.  \citet{GCN25484} later obtained SOAR/Goodman spectroscopy of this source and found that it was consistent with a Type IIb SN at $z=0.126$, which we confirm based on the same data.  Therefore, we rule out any association between GW190814 and AT~2019npw.

{\it AT~2019nqq}: AT~2019nqq \citep[also DESGW-190814c;][]{GCN25373} was discovered by the DESGW collaboration on 15 Aug.\ 2019.  We obtained a spectrum of this source on 17 Aug.\ 2019 \citep{GCN25379} that is dominated by a continuum with a broad emission line near 7000~\AA. We infer this line to be H$\alpha$, implying that AT~2019nqq is consistent with a Type II SN at $z \approx 0.07$, similar to the results reported by \citet{GCN25419}.  Therefore, we rule out any association between AT~2019nqq and GW190814.

{\it AT~2019nte}: AT~2019nte was detected by \citet{GCN25398} in DECam imaging at $i=20.95$~mag on 17 Aug.\ 2019.  We obtained a spectrum of this source with SOAR/Goodman as reported by \citet{GCN25784}. It had too little signal to obtain a robust spectroscopic classification, but we obtained a redshift of the underlying host galaxy from the Balmer features.  From this emission, we infer that AT~2019nte is at $z=0.0701\pm0.0006$.  At the location of AT~2019nte, this redshift is not ruled out by our volumetric cut, and so we consider AT~2019nte to be a viable candidate counterpart to GW190814 on the basis of its spectrum and redshift.

{\it AT~2019ntp}: The candidate AT~2019ntp was discovered by DECam/GROWTH \citep[also DG19gcwjc;][]{GCN25393} on 18 Aug.\ 2019 at $i=21.0$~mag.  We obtained a spectrum with SOAR/Goodman on 1 Sep. 2019 as reported by \citet{GCN25596}; AT~2019ntp is a Type Ia SN at $z=0.114$, and so we do not consider it to be associated with GW190814.

{\it AT~2019ntr}: Candidate transient AT~2019ntr was detected by DECam/GROWTH \citep[also DG19sbzkc;][]{GCN25393} on 18 Aug.\ 2019 at $z=21.2$~mag.  We obtained a SOAR/Goodman spectrum on 29 Aug.\ 2019 \citep{GCN25540} from which we determine that AT~2019ntr is a Type II SN at $z=0.218$.  Therefore, we do not consider AT~2019ntr to be associated with GW190814.

{\it AT~2019num}: The candidate AT~2019num was discovered by DECam/GROWTH on 18 Aug.\ 2019 at $i=21.3$~mag \citep{GCN25393}.  We obtained a spectrum with SOAR/Goodman on 27 Aug.\ 2019 \citep{GCN25484} from which we determine that AT~2019num is a Type II SN at $z=0.127$, and so it is not considered to be associated with GW190814.

{\it AT~2019omx}: The candidate AT~2019omx was discovered by the DESGW on 21 Aug.\ 2019 at $z=22.1$~mag \citep{GCN25486}.  We obtained a spectrum on 29 Aug.\ 2019 with SOAR/Goodman \citep{GCN25540}.  The spectral classification is inconclusive, although we are able to constrain the redshift from Balmer emission in the likely host galaxy as $z=0.164$.  Therefore, we rule out this candidate as being associated with GW190814 owing to its implied luminosity distance.

\textit{AT~2019osy} The GW190814 candidate EM counterpart AT~2019osy was discovered by the Australian Square Kilometre Array Pathfinder telescope (ASKAP) telescope \citep{2008ExA....22..151J, 2007PASA...24..174J,2016PASA...33...42M} as a rising radio source possibly in host galaxy 2dFGRS TGS211Z177 ($z=0.0738$) on 23 Aug.\ 2019 and reported to the community on 27 Aug.\ 2019 \citep{2019TNSTR1637....1S}.  We obtained a long slit Keck/DEIMOS spectrum of AT~2019osy and were unable to identify a transient. The extract spectrum is best matched by a galaxy template in SNID and we therefore determine that the transient is not detected.

\subsection{Premerger Variability toward Candidate Counterparts}\label{sec:variability}

We analyzed the Pan-STARRS DR2 Detection catalog, ASAS-SN Photometry database, and Catalina Sky Survey DR2 catalog for coincidence with each of our candidates as described above.  The PS1 DR2 Detection catalog separates multi-epoch observations into separate observations.  We found two $g$-band detections coincident with AT~2019nto on 8 Sep. 2010 with an aperture flux of $g=20.08\pm0.02$~mag and on 9 Oct. 2010 at $g=21.56\pm0.05$~mag, indicating that the source is variable. It corresponds to the likely host galaxy WISEA~J004203.40-244820.4 \citep{Chung+11}, with a nominal offset of 0.28\arcsec\ between AT~2019nto and the host galaxy.  Therefore, we consider AT~2019nto to be unassociated with GW190814.

As described by \citet{Jayasinghe19}, the ASAS-SN photometry catalog was initially constructed from the $>$50 million point sources with $V<17$~mag in the APASS catalog \citep{Henden15}.  Thus, sources with counterparts in the ASAS-SN catalog are not necessarily variable, but we can use the catalog to crossmatch to candidate variables and check for multiple premerger detections.  We identified a single candidate AT~2019nup that had a counterpart in the ASAS-SN photometry database\footnote{This source is called AP16326416 in the ASAS-SN Photometry Database.}.  This source is coincident with the center of a $z=0.03665\pm0.00014$ galaxy in the Southern Abell Redshift Survey called SARS~013.16023-27.04103 \citep{Way05}, with a 0.71\arcsec\ between AT~2019nup and its host.  The AT~2019nup premerger counterpart exhibited a previous outburst on 15 May 2016 at $V=15.85$~mag, and so we rule out any association with GW190814.

Despite these findings, it is possible that AT~2019nto or AT~2019nup are associated with AGN activity as a result of a compact object merger \citep[similar to][]{Graham20}.  We revisit these objects in \autoref{sec:agn}.

\subsection{Redshifts of Candidate Host Galaxies}\label{sec:specz}

We associate candidates with their likely host galaxies on the basis of projected separation from galaxies with spectroscopic redshifts in the NASA/IPAC Extragalactic Database (NED)\footnote{\url{http://ned.ipac.caltech.edu/}}.  We require that the host-transient separation is at most $20'$ to be considered a candidate host galaxy and 300~kpc assuming the source is located at the host redshift and calculating an angular diameter distance using \citet{Planck15} cosmology.  These criteria ensure that nearby extragalactic transients are selected in galaxies at large projected separations --- notably AT~2019npd and AT~2019nvb, low-luminosity transients in the outer arms of the Sculptor Galaxy and separated by $9.8'$ (10.0~kpc) and $12.6'$ (12.7~kpc), respectively \citep{GCN25362,GCN25417}.  If there are multiple candidate host galaxies, we choose the galaxy with the smallest transient-host separation (in kpc).  All of our host-galaxy redshifts are noted in \autoref{tab:candidates}.

In addition, we measured redshifts to host galaxies of candidates without spectroscopic redshifts in NED with Shane/Kast and SOAR/Goodman (\autoref{tab:spectra}; note that 2 objects, AT~2019nra and AT~2019nte, in our table were observed twice and 8 other objects has spectra that were too low quality to measure redshifts).  From these spectra we infer the spectroscopic redshift to the candidate host galaxies. Spectroscopic redshifts were measured by cross-correlating spectra with galaxy templates using the RVSAO package \citep{Kurtz98}.  In addition to visual inspection, the \citet{Tonry79} cross-correlation height-to-noise ratio ($r$) was used to determine the quality of the redshift match for each template, with $r > 4$ as our threshold for a reliable redshift.

The plurality of candidates we rule out are due to host-galaxy associations outside the 99th percentile volume provided by the LVC, totaling \Nspecz\ with spectroscopic redshifts and \Nphotoz\ with photometric redshifts.  Given the selection criteria and chance coincidence calculation described above, we infer that these associations are robust and it is unlikely that we have associated these transients with foreground or background galaxies outside the LVC volume by chance.

However, given the large overall fraction of galaxies with photometric redshifts and the statistical and systematic uncertainties associated with photometric redshift catalogs \citep[see][]{Bilicki13,PS1redshift,Zhou20}, it is possible that some host galaxies are indeed in the LVC volume.  We avoid this possibility by requiring that, in order to be considered ``outside the LVC volume,'' the photometric redshift $z_{\rm phot}$ and uncertainty $\sigma_{z_{\rm phot}}$ must satisfy $z_{\rm phot} + \sigma_{z_{\rm phot}} < z_{\rm LVC} - \sigma_{z_{\rm LVC}}$ or $z_{\rm phot} - \sigma_{z_{\rm phot}} > z_{\rm LVC} + \sigma_{z_{\rm LVC}}$ (where $z_{\rm LVC}$ and $\sigma_{z_{\rm LVC}}$ are the redshift and uncertainty implied by the luminosity distance the LVC HEALPix map and assuming \citet{Planck15} cosmology).  For analyses using GW190814 to measure the Hubble constant statistically \citep[as  by][]{Soares-Santos19,LIGO-O2-cosmology,Palmese20}, additional spectroscopic observations of galaxies in the GW190814 volume will be essential to reduce the uncertainties in host-galaxy associations and the Hubble constant.

Some apparently ``hostless'' sGRB afterglows have been localized to regions of the sky with no apparent host galaxy to deep limits of $V$$\approx$26~mag \citep{Fong13,Behroozi14}.  This finding suggests that some BNS and NSBH binaries experience natal kicks of $>$150~km~s$^{-1}$ and potentially travel hundreds of kpc before merging and producing GW signals detectable by the LVC.  This scenario is more likely in the local Universe where merging NS systems are more heavily weighted toward long delay times and thus systems that have travelled further from their birth environments \citep{Anand17,Safarzadeh19}.  We do not consider apparently hostless transients to be ruled out as candidate counterparts to GW190814, and so the primary effect of a large projected separation between a transient and host on our analysis is the increased probability of a chance coincidence.  A more detailed analysis would marginalize over the likelihoods of all candidate host galaxies for all candidate counterparts \citep[as in, e.g.,][]{Aggarwal21} and only cut candidates if they are associated with a galaxy outside the localization volume with high probability.

\subsection{Photometric Evolution of Candidates}\label{sec:candlightcurve}

To constrain the optical properties of each candidate, we assign them luminosity distances on the basis of redshift (where available in \autoref{tab:candidates}) or the luminosity distance at the transient location in the LVC GW190814 HEALPix map.  Thus, we infer the absolute magnitude of every candidate that is not a minor planet in our analysis.  Many of these sources are inferred to have extremely luminous absolute magnitudes ($<-21$~mag), likely reflecting interloping AGNs found in galaxy-targeted searches \citep[e.g.,][]{VeronCetty10,Smith20}, and so most of these sources were ruled out on the basis of host-galaxy associations and our volume cut.

\begin{figure*}
    \centering
    \includegraphics[width=0.95\textwidth]{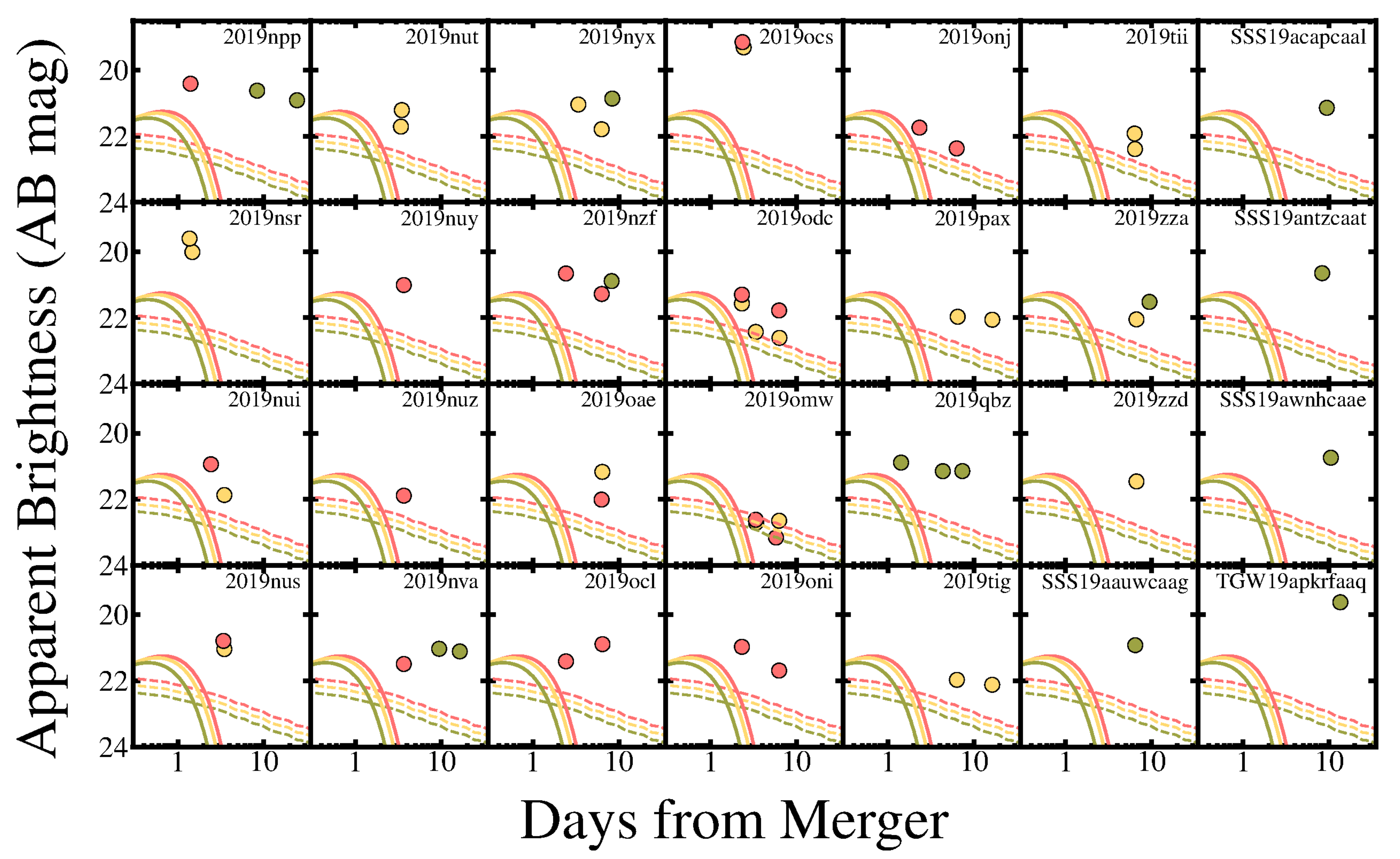}
    \caption{$riz$ (green, yellow, and red, respectively) light curves of candidate counterparts to GW190814 that we ruled out on the basis of photometric cuts as described in \autoref{sec:candlightcurve}.  We show photometry from our own follow up (\autoref{tab:cand-photometry}), DECam \citep{Andreoni19b,Morgan20}, and VLT-GRAWITA \citep{Ackley20} for each source that we rule out on the basis of light curve parameters.  For comparison, we show blue kilonova light curves (solid lines) and $\gamma$-ray burst afterglow light curves corresponding to the off-axis GRB~170817A analog (dashed lines) corresponding to the models shown in \autoref{fig:lightcurves}.  We place each model light curve at the distance of each candidate using redshift information where available or the luminosity distance at its location in the GW190814 localization map \citep[from][]{Abbott20}.  In general, the detected candidates are significantly more luminous than the models we consider and at the observed epochs, which enable us to rule out these candidates as too luminous to be a counterpart to GW190814 in \autoref{tab:cand-analysis}.}
    \label{fig:cand-lightcurves}
\end{figure*}

Guided by theoretical models for kilonovae, we note that there are no plausible models more luminous than $-18.0$~mag \citep[e.g., in][]{Metzger16,Kasen17,Metzger19,Morgan20}, and thus use this limit to exclude any particularly luminous sources comprising 14 candidates.  In addition, where available we analyzed any variability in all photometric bands and rule out sources that decline more slowly than $0.1$~mag~day$^{-1}$ over a baseline of at least 3~days.  We rule out 6 candidates (AT~2019omw, AT~2019qbz, AT~2019zza, 2019aako, 2019aakp, 2019aakq) as candidates that were either rising or slowly declining at the time of observation.  In addition, we do not detect some candidates in our follow-up program after they were reported by previous groups.  Thus, we can place a lower limit on their decline rate assuming the source was fainter than the limit at the time of observation \citep[in particular, for limits reported by][]{Andreoni19b,Morgan20}.  We rule out 8 candidates based on this criterion.  The light curves and limits for all candidates ruled out by photometric cuts are shown in \autoref{fig:cand-lightcurves}.

We acknowledge that these cuts are model-dependent, and so the \Nphot\ candidates rejected at this stage remain viable if there is a potential NSBH EM counterpart model that matches any of the criteria on which we cut.  In particular, we directly compare our empirical limits on the minimum absolute magnitude and decline rate (\autoref{sec:linear}) to these candidates in order to highlight that our limits are consistent with sources discovered by our surveys and in the literature.  However, our photometric constraints are the weakest criteria for ruling out candidate counterparts and only pertain to candidates that cannot be ruled out in any other context.

\section{Limits on Electromagnetic Counterparts}\label{sec:limits}

Having ruled out all known candidate optical transients as viable counterparts to GW190814, we estimate the depth of all of our imaging and show our results for every individual image in \autoref{tab:observations}.  In the remainder of this section, we then analyze these limits in the context of model optical counterparts to GW190814 and place constraints on the properties of any hypothetical counterpart.

\subsection{Estimates of the Limiting Magnitude}\label{sec:fakestars}

\subsubsection{1M2H}

We estimate limiting magnitudes for Nickel, Swope, and Thacher difference images by planting artificial sources in each individual survey image, allowing us to determine the probability of finding a source as a function of magnitude.  This procedure requires the following steps.

\begin{enumerate}

    \item {\it Building the PSF model}: The PSF parameters for each image are determined by {\tt DoPhot} \citep{Schechter93}.  We then rescale the PSF model such that the magnitude determined from the sum of the pixels in the model (given the predetermined zeropoint of the image) is equal to the magnitude we wish to simulate.

    \item {\it Planting the artificial sources}: For each image, we plant 1500 artificial sources in the regular, unsubtracted survey image.  These sources are randomly placed across the image, with magnitudes randomly chosen from a flat distribution between 18 and 25~mag.

    \item {\it Running the reduction pipeline}: Once the sources have been planted, we run the difference-imaging and source-detection pipeline with the exact same pipeline stages used for transient discovery, beginning at the difference-image stage to incorporate correlated pixel noise and subtraction artifacts into the detection-efficiency calculation.  We have verified that using 1500 sources does not adversely affect the quality of the difference images.

    \item {\it Measuring the detection efficiency}: For bins of 0.2 mag, we compute the detection efficiency by dividing the fraction of sources detected in the difference image by the total number of simulated sources within that magnitude bin.  This gives us the full detection efficiency as a function of magnitude, as well as the magnitude at which 50\% of artificial sources are recovered.  We define the 3$\sigma$ limiting magnitude by interpolating our efficiency curves to the magnitude at which 99.7\% of the average maximum fraction of recovered sources at any magnitude are at least as bright as that magnitude threshold.  This fraction can be less than 100\% if sources land on cosmic rays or on top of very bright stars, and we reweight our detection efficiency for that image by the maximum recovered fraction.  These values correspond to the limiting magnitudes given for Nickel, Swope, and Thacher images in \autoref{tab:observations}.

\end{enumerate}

These steps allow an ``end-to-end" measurement of image-detection efficiency that includes correlated noise from the difference image convolution procedure and the ability of our automated software to recover point sources in our images.  Correlated difference image noise in particular would not be measurable from a simple sky-background-based computation.  These limits are also consistent with our cuts in transient identification, which are based on the same pipeline and identification process.

As discussed above, the limiting magnitude in KAIT and Las Cumbres images was estimated using the 3$\sigma$ scatter in the background measured across each image.

\subsection{Constraints on EM Counterparts to GW190814}\label{sec:constraints}

We have demonstrated that there is no candidate counterpart to GW190814 in \autoref{sec:candidates}.  In the following analysis, we determine the joint limits on the presence of an EM counterpart to GW190814 using all difference imaging (neglecting our template images and Keck/MOSFIRE images for which we do not have templates).  In practice, this involves comparison between the expected in-band light curves for various transient sources and the 3$\sigma$ limiting magnitudes as a function of sky position and time relative to merger reported in \autoref{tab:observations}.  In order to provide a physically meaningful limit, we emphasize that our uniform sample of difference images is reduced self-consistently with (1) template exposures of each targeted field obtained in the same filter and instrumental configuration, (2) true PSF-convolved difference imaging between all science exposures and the template exposures as described in \autoref{sec:obs}, and (3) limiting magnitudes derived in the difference images themselves and consistent with the signal-to-noise ratio of any detected transient sources.  In this way, we are confident that our limiting magnitudes rule out any EM counterparts across our imaging to the magnitude level quoted in \autoref{tab:observations}.  We emphasize that while our limiting magnitudes account for Milky Way extinction in the direction of each image, we assume there is no additional source of extinction, for example in the host galaxy or local environment of the GW190814 progenitor system.

We obtained the latest GW190814 HEALPix\footnote{\url{http://healpix.sourceforge.net}} \citep{HEALpix} sky-localization map from the Gravitational-Wave Candidate Event Database\footnote{\url{https://gracedb.ligo.org/}} and presented by \citet{Abbott20}.
All observations and limiting magnitudes reported in \autoref{tab:observations} were then ingested into our GW planning and analysis code {\tt teglon} (see \autoref{sec:obs}) and crossmatched to the corresponding HEALPix pixel elements. At each pixel, we estimated the mean and standard deviation of the best-fitting LVC distance using the {\tt moments\char`_to\char`_parameters} from the {\tt ligo.skymap.distance} Python package \citep{GTD, GTDSupplement}. Each resulting Gaussian is then truncated at zero distance and renormalized such that the total three-dimensional probability (i.e., integrated over all sky pixels and luminosity distances) is unity.

We then consider the likelihood that an optical counterpart to GW190814 would be detected for a given counterpart model, which we generically classify as kilonovae (\autoref{sec:kilonova}), sGRBs (\autoref{sec:sgrb}), and linearly-evolving optical counterparts (\autoref{sec:linear}).  Each model provides an estimate of the brightness of the counterpart as a function of time, wavelength, and distance, which we transform to in-band light curves using {\tt pysynphot}.

Over the GW190814 localization region, most areas were observed with multiple images and in multiple filters. Therefore, to compute the net detection efficiency given all sky map pixels and all observations, we determine the detection efficiency in a given model for each sky map pixel across each epoch and filter, and then combine the result into a cumulative detection efficiency. To do this, we retrieve the absolute magnitude for a given model at the time and in the band of each observation. Our inferred detection efficiency is also dependent on specific image- and line-of-sight dependent quantities, for example the Milky Way extinction and limiting magnitude as described in \autoref{sec:fakestars}.

We then reparameterize our limiting magnitude in terms of the distance $D_{\mathrm{model}_{j,f}}$ at which we would expect to detect a source in image $j$ with a filter $f$ and a limiting magnitude $m_{j,f}$, where the source has an absolute magnitude $M_{\mathrm{model}_{j,f}}$ and line-of-sight extinction $A_{f}$, as

\begin{eqnarray}
\mu_{\mathrm{model}_{j,f}} =  m_{j,f} - M_{\mathrm{model}_{j,f}} - \mathrm{A}_{f} \\
D_{\mathrm{model}_{j,f}}~\mathrm{[Mpc]} = 10^{0.2 \times (\mu_{\mathrm{model}_{j,f}} - 25)}.
\end{eqnarray}

\noindent For each pixel $i$ in the LVC localization map that overlaps with image $j$, we then calculate the relative efficiency of detecting this model by integrating the pixel distance distribution from zero distance to $D_{\mathrm{model}_{j,f}}$,

\begin{equation}
P_{\mathrm{model}_{i,j}} = \frac{1}{\sqrt{2\pi} \sigma_{D_i}} \int_0^{D_{\mathrm{model}_{j,f}}} e^{-\frac{1}{2} \left(\frac{\bar{D_{i}} - D}{\sigma_{D_i}}\right)} dD,
\end{equation}

\noindent where $\bar{D_{i}}$ is the mean distance and $\sigma_{D_{i}}$ is its standard deviation corresponding to the pixel $i$ in the localization map.

To combine independent observations that overlap with each pixel, we take the complement of the joint probability that we do not see the source in any epoch. That is, for each pixel, we weight the relative likelihood that we would detect a specific model in image $j$ by the LVC 2D pixel probability in each pixel $P_{i}$ and sum over all pixels to obtain a cumulative probability of detecting a specific model,

\begin{equation}
P_{\mathrm{model}} = \displaystyle\sum_{i}  P_{i} \left[1 - \prod_{j} \left(1 - P_{\mathrm{model}_{i,j}}\right)\right].
\end{equation}

\noindent This final probability, which we interpret as the likelihood that we would have seen a source with properties described by a model for the EM counterpart to GW190814, is calculated for a wide range of models described below.

\subsection{Limits on Kilonovae}\label{sec:kilonova}

The counterpart to GW170817 was initially localized by targeting optical emission from a kilonova \citep{Abbott17:detection,Coulter17,Kasen17}, or a transient powered by the decay of $r$-process species synthesized in ejecta from a NS merger.  As in \citet{Kilpatrick17}, we parameterize this source for a given ejecta mass $m_{\rm ej}$ and velocity $v_{\rm ej}$.  Following the numerical model presented by \citet{Metzger16} and used by \citet{Coughlin19,Coughlin20b}, we allow for a varying electron fraction $Y_{e}$ in the ejecta, which affects the neutronization and thus the composition of the ejecta.  Broadly speaking, low electron fractions will lead to a high level of neutronization and thus heavier $r$-process species, which tend to be more optically thick at the optical and near-infrared wavelengths of our search.  The electron fraction and distribution of $r$-process species have a secondary effect on the radioactive heating rate \citep[e.g., see][]{Lippuner15}, which is incorporated into our light curves, but the ejecta opacity is the dominant effect of varying $Y_{e}$.

In order to accurately estimate the ejecta opacity as a function of electron fraction, we adopt the mean opacity for kilonovae with varying compositions presented by \citet[][see their Table 1]{Tanaka19} but with a floor of $\kappa=1.0~\mathrm{cm}^{2}~\mathrm{g}^{-1}$ for $Y_{e}>0.40$.  This results in relatively high opacities of $\kappa>30.0~\mathrm{cm}^{2}~\mathrm{g}^{-1}$ for $Y_{e}\lesssim0.15$, and so this model is somewhat pessimistic compared with similar treatments by \citet{Andreoni19} and \citet{Coughlin20b}.  However, we are confident that our models accurately reflect a broad range of ejecta composition for varying masses and velocities.

\begin{figure*}
    \centering
    \includegraphics[resolution=300,width=0.49\textwidth]{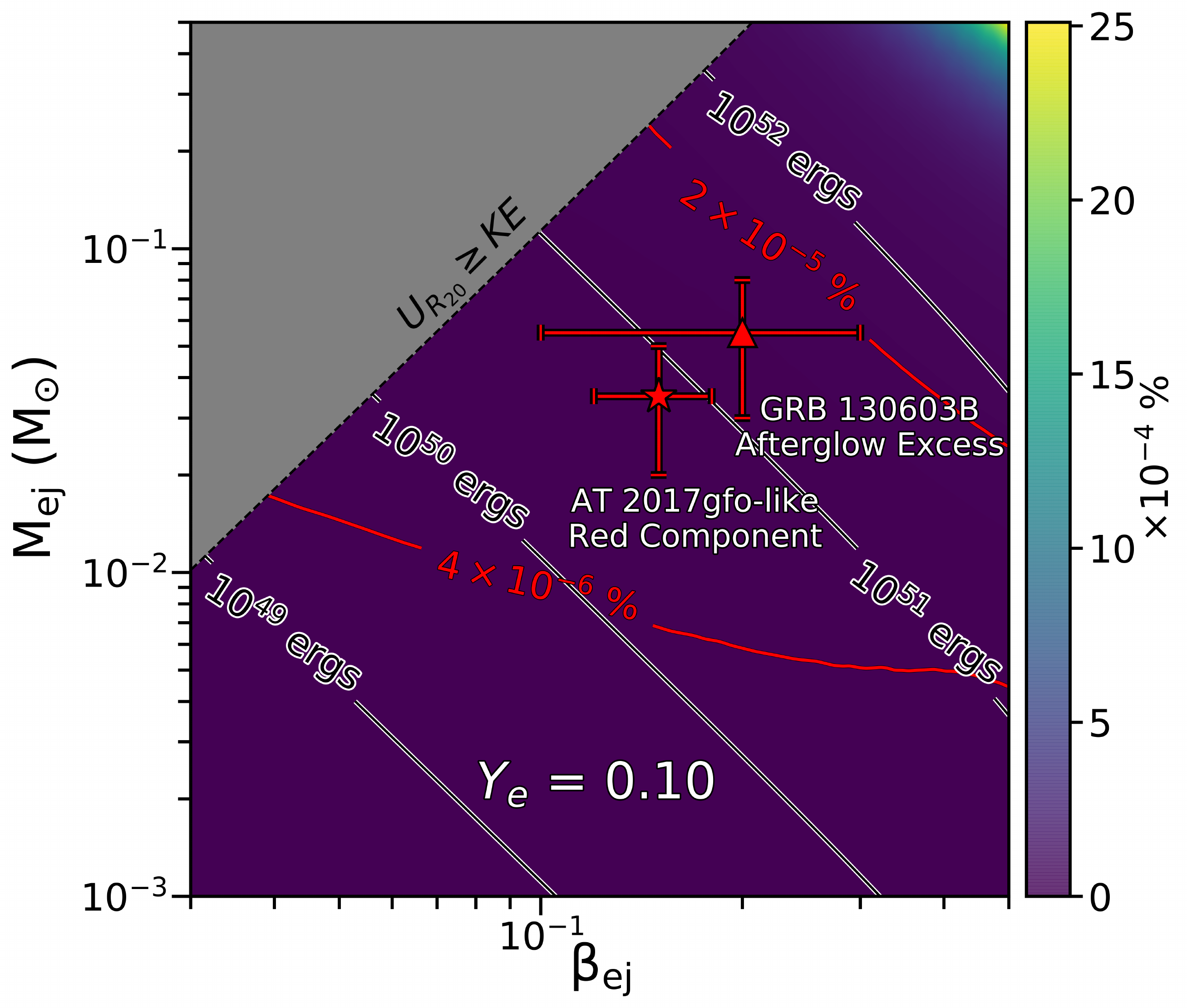}
    \includegraphics[resolution=300,width=0.49\textwidth]{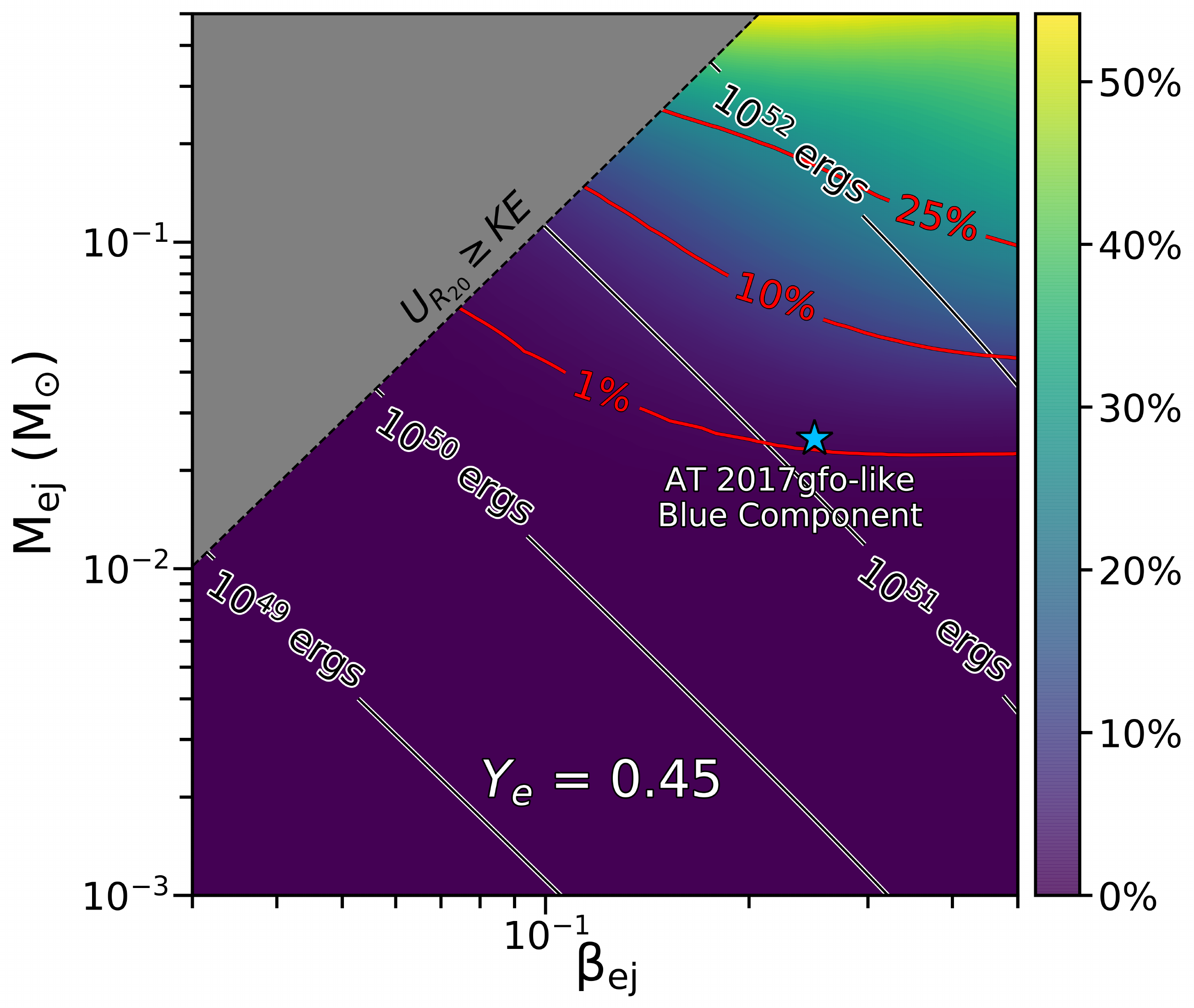}
    \caption{Constraints on the presence of a kilonova assuming an electron fraction corresponding to a ``red'' kilonova ($Y_{e}=0.10$, {\it left}) and a ``blue'' kilonova ($Y_{e}=0.45$, {\it right}).  For both sets of models, we show the estimated likelihood that we would have detected a source for a given ejecta mass ($m_{\rm ej}$ in \Mdot) and velocity ($\beta_{\rm ej}$ in natural units) following the procedure in \autoref{sec:constraints}.  We have grayed out the region where the binding energy of the ejecta (assuming an extremely stiff neutron star radius of $20$~km) exceeds its kinetic energy.  For context, we show contours of equal probability in red and lines of equivalent ejecta kinetic energy in black.  Finally, in each panel we show the location of the corresponding kilonova ejecta component AT~2017gfo as derived by \citet{Kilpatrick17} (specifically $\beta_{\rm ej}=0.15$, $M_{\rm ej}=0.035~M_{\odot}$ for the red kilonova and $\beta_{\rm ej}=0.25$, $M_{\rm ej}=0.025~M_{\odot}$ for the blue kilonova) as well as the putative kilonova counterpart to GRB130603B as described by \citet{Tanvir13}.}
    \label{fig:kilonova}
\end{figure*}

We estimated the in-band light curves for ejecta masses 0.001--0.5~$M_{\odot}$ and velocities (0.001--0.5)c.  We show our estimated probability of detection for our fiducial models, a ``red'' kilonova ($Y_{e}=0.10$) and a ``blue'' kilonova ($Y_{e}=0.45$), in \autoref{fig:kilonova}.

\subsection{Limits on sGRBs}\label{sec:sgrb}

We adopted the models of \citet{Duffell13} and \citet{Wu18} to model potential sGRB optical counterparts to GW190814.  We parameterize these models by the isotropic kinetic energy $E_{\rm K,iso}$ and the circumburst density $n$ as well as the viewing angle of the sGRB jet $\theta_{\rm obs}$ (assuming a jet opening angle identical to GRB170817A of $\theta_{0}=5.2^{\circ}$).
In addition to these variables, we assume that the specific internal energy $\eta_{0}=7.9$,  boost Lorentz factor $\gamma_{B}=9.4$, spectral index $p=2.15$, electron energy fraction $\epsilon=0.1$, and magnetic energy fraction $\epsilon_{B}=2.5\times10^{-4}$ from the updated GRB170817A analysis by \citet{Wu19}.

We model a range of isotropic kinetic energies (10$^{48}$ to 10$^{52}$~erg) and circumburst densities (10$^{-6}$ to 1~cm$^{-3}$), roughly spanning the range of sGRB jet parameters presented by \citet{Fong15}.  In addition, we model two jet viewing angles --- an on-axis model with $\theta_{\rm obs}=0^{\circ}$ and an off-axis model with $\theta_{\rm obs}=17^{\circ}$ (see characteristic light curves in \autoref{fig:lightcurves}).  In the latter case, the afterglow light curve is still rising after our latest observations of the GW190814 field, and so our limits our primarily sensitive to the luminosity of the afterglow during our last observation epoch, corresponding to Swope and Nickel observations obtained on 10--11 Sep.\ 2019 with $3\sigma$ limiting magnitudes around $r=20.0$--21.5~mag.  Our limits are mostly insensitive to sGRB afterglow light curves with jet viewing angles greater than $17^{\circ}$ assuming a jet opening angle of $\theta_{0}=5.2^{\circ}$ as in \citet{Wu18}.

\begin{figure*}
    \centering
    \includegraphics[resolution=300,width=0.49\textwidth]{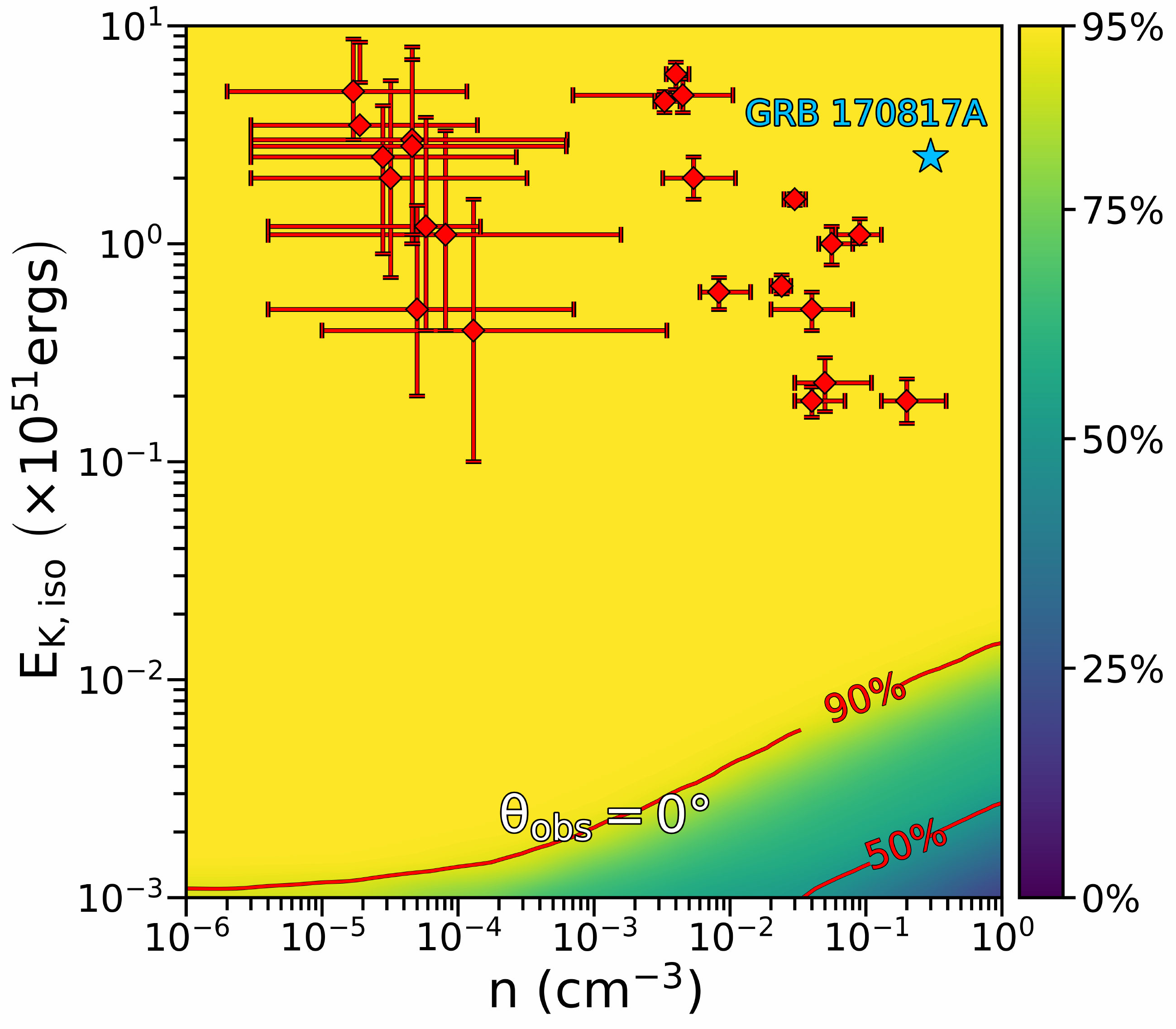}
    \includegraphics[resolution=300,width=0.49\textwidth]{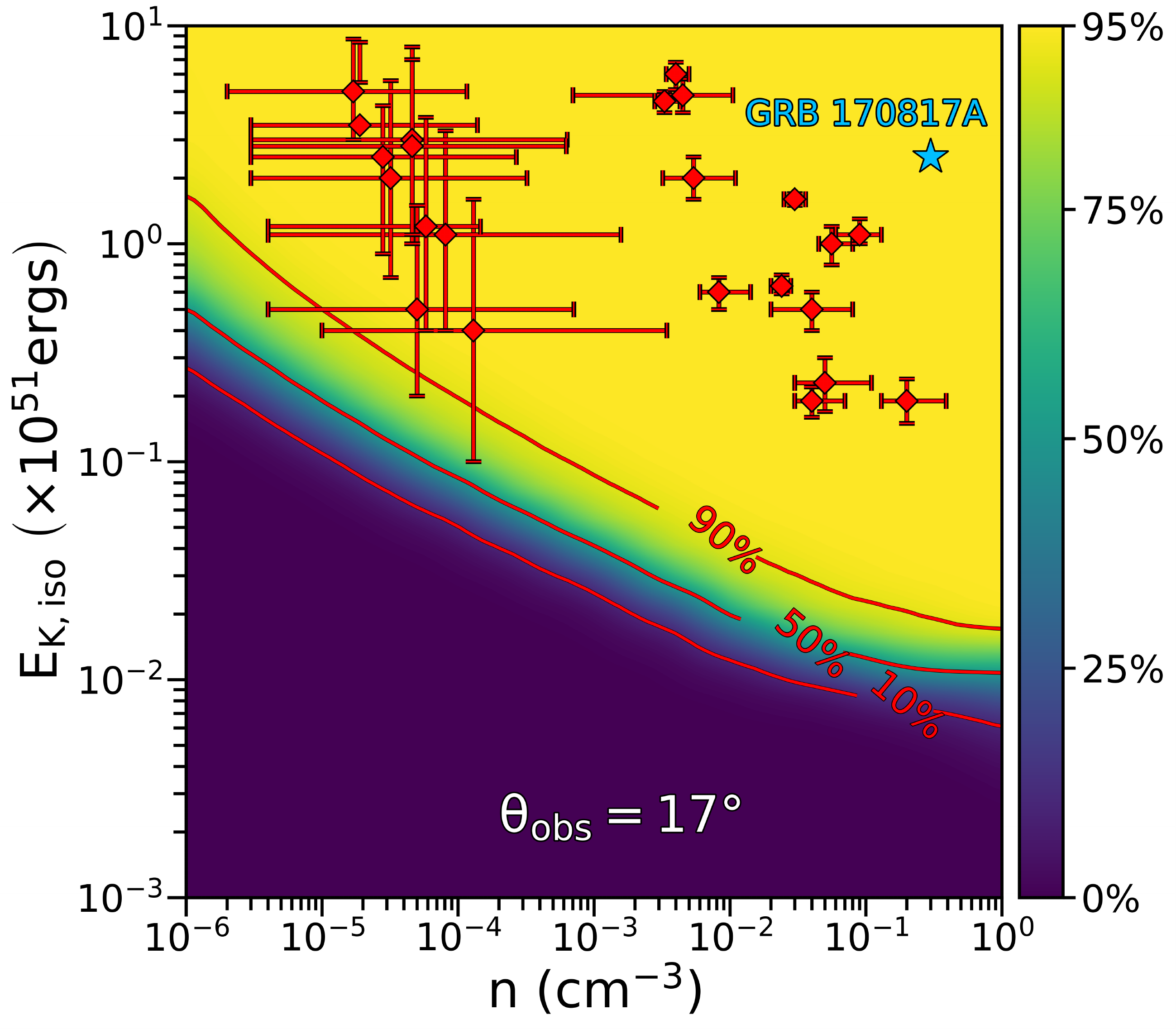}
    \caption{Constraints on the presence of an sGRB viewed on-axis ({\it left}, $\theta_{\rm obs}=0^\circ$) and off-axis ({\it right}, $\theta_{\rm obs}=17^{\circ}$).  In both cases, we show the total observed two-dimensional LVC probability weighted by the likelihood that we would observe a counterpart with a specific explosion energy ($E$ in units of $10^{51}~\mathrm{erg s}^{-1}$) and circumburst density ($n$ in units of $\mathrm{cm}^{-3}$).}
    \label{fig:sgrb}
\end{figure*}

\subsection{Generic Limits on EM Counterparts}\label{sec:linear}

The final set of models we consider is defined empirically using a peak absolute magnitude and linear decline rate in units of mag~day$^{-1}$.  As kilonovae and sGRBs are rapidly evolving with extremely short rise times \citep[i.e., for kilonovae with physical parameters similar to those of AT~2017gfo and sGRBs viewed on-axis;][]{Arcavi17,Drout17,Kilpatrick17}, we model these generic EM counterparts with a peak brightness at the time of merger as defined by the GW signal.  In order to combine our limits across all wavelengths used to follow up GW190814, we further assume that the model counterpart has a flat spectral energy distribution such that the source has the same magnitude in all bands.

\begin{figure}
    \centering
    \includegraphics[resolution=300,width=0.49\textwidth]{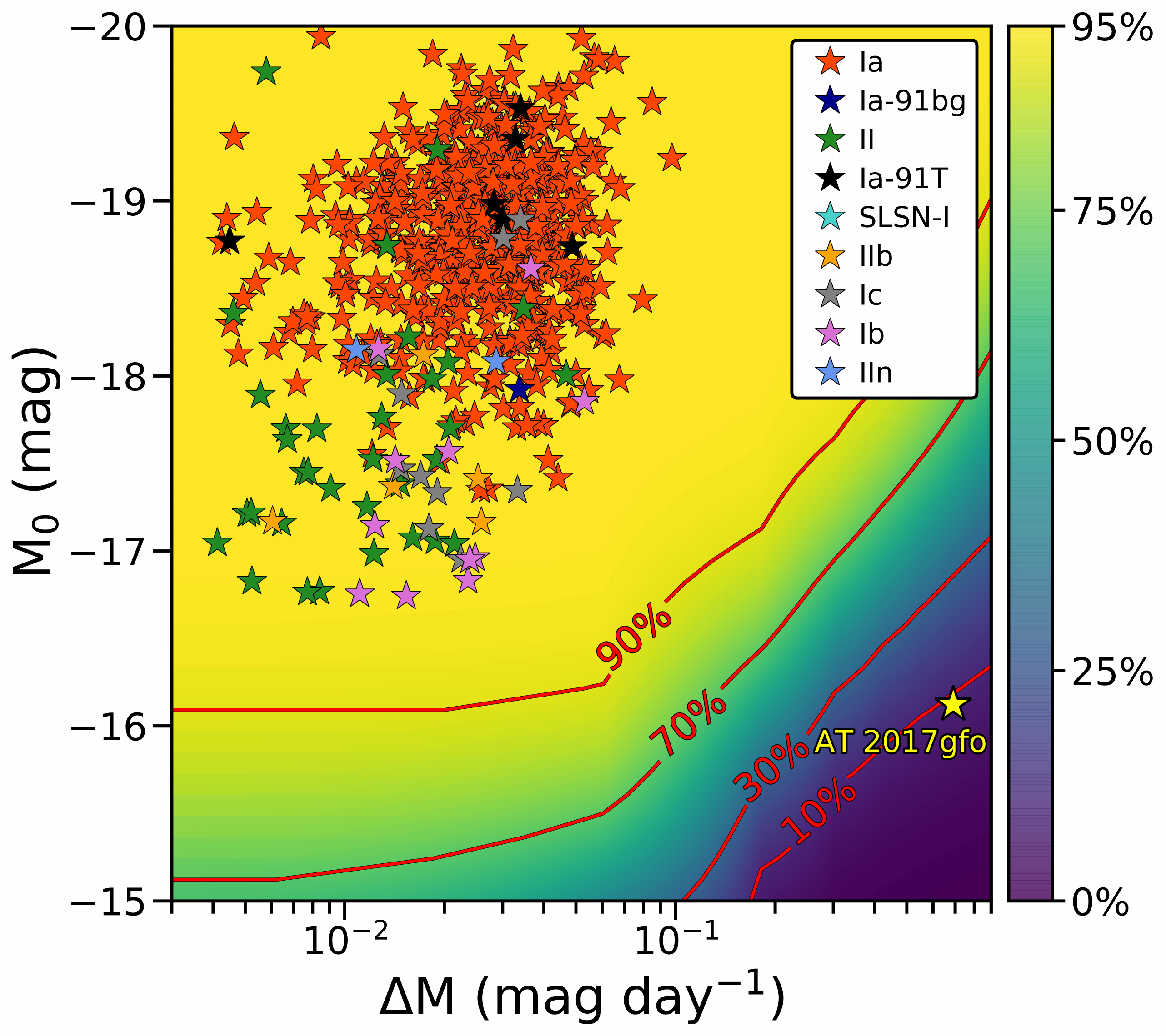}
    \caption{Constraints on the presence of a generic optical counterpart assuming peak magnitude $M_{0}$ occurs approximately at the time of merger ($t_{0}$) and the source declines at a linear rate $\Delta M$ in $\mathrm{mag day}^{-1}$.  We further assume that the counterpart has a flat spectral shape at all times such that its magnitude is $M_{0} + (t - t_{0})\,\Delta M$ in all bands for all times $t$.  We show the properties of AT~2017gfo with a yellow star with $r$-band peak $-$16.5~mag and decline rate 0.3~mag~day$^{-1}$ \citep[based on][]{Siebert17,Drout17}.  For comparison, we show the peak magnitudes and decline rates of a range of other optical transients derived from \citet{Siebert17}.  We rule out a AT~2017gfo-like counterpart (yellow star) with 6.0\% confidence.}
    \label{fig:linear}
\end{figure}

Following the procedure outlined above, we combine all of our limits into a likelihood of detection assuming a uniform range of peak absolute magnitudes (from $-15$ to $-20$~mag) and decline rates (from $3\times10^{-3}$ to 1~mag~day$^{-1}$ in log space) as shown in \autoref{fig:linear}.  For comparison, we show our joint limits along with the peak magnitudes and decline rates of various astrophysical transient sources \citep[taken from][]{Siebert17}.

Although our limiting magnitudes are relatively strong compared with SNe and other luminous transient sources, our empirical limits only extend to -16~mag ($\approx$21~mag at 240~Mpc) for slowly-evolving (declining at $<$0.1~mag~day$^{-1}$) events given that we observed the GW190814 localization over 3--4~days.  This places AT~2017gfo-like transients outside of what we can detect, as it began to decline at $>$0.3~mag~day$^{-1}$ in blueward of $i$-band at less than 1~day from merger \citep{Drout17}.

\section{Discussion}\label{sec:discussion}

In this section, we describe the physical implications of our limits in the context of candidate EM counterpart models described in \autoref{sec:limits}.  As the total area covered by our observations comprises $\sim$95\% of the total two-dimensional probability in the latest GW190814 map, the strongest constraints on EM counterparts are at this significance level.  We then discuss further implications of these limits in the context of likely merger models, incorporating the gravitational-wave data from \citet{Abbott20}.  Finally, in \autoref{sec:agn} we consider the scenario where GW190814 occurred in an AGN disk.

\subsection{Joint Limits on Electromagnetic Counterparts}

For an AT~2017gfo-like blue kilonova as described in \autoref{sec:constraints}, we estimate a 50\% chance of detecting the counterpart at distances $<$95~Mpc.  However, the overall probability of detection given the best-fitting LVC distances for this source is around 1\%.  As shown in \autoref{fig:lightcurves}, these limits are dominated by our early-time observations, as the kilonova models decline rapidly overall.  Similarly, the probability of detecting an AT~2017gfo-like red kilonova is extremely low at $\sim${}$4 \times 10^{-7}$\%, which is also mainly constrained by our early-time Swope $r$-band and Las Cumbres $i$-band observations.  We are only sensitive to red kilonovae under the assumption that the true distance to GW190814 is much closer ($<$80~Mpc) than the reported LVC best-fitting distance.  Finally, if the counterpart had colors between the blue and red model but the same overall luminosity as AT~2017gfo, the recovery fraction would likely be between the red-only and blue-only numbers.

Inverting these constraints, we are sensitive to the blue kilonova models with $\beta_{\rm ej}=0.25$ and $M_{\rm ej}>0.5~M_{\odot}$ at 30\% significance or $M_{\rm ej}>0.3~M_{\odot}$ at 25\% significance.  These limits are extreme for GW190814, especially considering that even the most optimistic merger models for a 2.59$\pm$0.08~$M_{\odot}$ and 23.2$\substack{+1.0\\-0.9}$~$M_{\odot}$ NSBH merger predict effectively no ejecta mass.

Our limits for GRB models are significantly more constraining given plausible counterparts and small viewing angles.  In particular, we find that all on-axis models are ruled out at our maximum 95\% likelihood except for the most extreme low-energy ($\sim${}$10^{48}$~erg) bursts.  In particular, we would have seen the on-axis optical afterglow of a burst similar to any of the bursts described by \citet{Fong15} or a GRB170817A-like on-axis optical afterglow \citep[following $n=0.3$~cm$^{-3}$ and $E_{k}=2.5\times10^{51}$~erg in][]{Murguia-Berthier17}.  Varying the physical parameters of these on-axis, GRB170817-like bursts in \autoref{fig:sgrb}, we can rule out bursts with circumburst density $n=0.3$~cm$^{-3}$ and isotropic equivalent energy $E_{k}=2.5\times10^{51}$~erg at 95\% significance, $n=0.3$ and $E_{k}=2.5\times10^{48}$~erg at 50\% significance, and $n=10^{-6}$~cm$^{-3}$ and $E_{k}=2.5\times10^{51}$~erg at 95\% significance.  These limits are consistent with the non-detection of GRBs by INTEGRAL, which observed the localization region of GW190814 117$^{\circ}$ off axis and constrained the isotropic equivalent energy of the burst to be $<2.1\times10^{48}$~erg for a short GRB spectrum with spectral index $\alpha=-0.5$ \citep[assuming $D_{L}=239$~Mpc;][]{GCN25323}.  Similar measurements from just a few sources can lead to statistical constraints on jet geometries \citep{Farah20}.

For larger jet viewing angles ($\theta_{\rm obs}$), but a fixed jet opening angle of $\theta_{0}=5.2^{\circ}$, we are increasingly less sensitive to optical afterglows, primarily because the optical luminosity is significantly lower at early times where the majority of our limits are.  Beyond $\theta_{\rm obs}>17^{\circ}$, we are no longer able to rule out optical afterglows from GRBs similar to those of \citet{Fong15}.  In physical terms, we can rule out off-axis optical afterglows from bursts with $n=0.3$~cm$^{-3}$ and $E_{k}=2.5\times10^{51}$~erg at 95\% significance, $n=0.3$ and $E_{k}=2.5\times10^{49}$~erg at 50\% significance, and $n=10^{-6}$~cm$^{-3}$ and $E_{k}=2.5\times10^{51}$~erg at 62\% significance.  GW190814 had an inclination angle constrained from its gravitational wave signal of 40--70$^{\circ}$ \citep{Abbott20} implying that any associated afterglow was likely to be even further off axis and thus undetectable by our follow up.

Finally, our limits are comparable to the luminosities of most SNe and optical transients similar to those discussed as interlopers to potential GW counterparts by \citet{Siebert17}.  As shown in \autoref{fig:linear}, while we would be able to detect or rule out the presence of virtually all SN subtypes at 241~Mpc, we would likely not be able to detect an AT~2017gfo-like transient (ruled out at 6.0\% significance), which is consistent with kilonova model limits.

Assuming a faint counterpart with $\Delta M=0.68$~mag~day$^{-1}$ \citep[similar to the initial $r$-band decline rate for AT~2017gfo;][]{Drout17,Siebert17}, we rule out sources with $M_{\mathrm{peak}}<-17.8$~mag at 50\% significance.  Similarly, if we assume that the initial magnitude of the source at the time of merger is similar to that of AT~2017gfo with $M_{r}=-16.1$~mag, we can rule out sources that decline with $\Delta M=0.06$~mag~day$^{-1}$ at 50\% significance.

\subsection{Combining GW and EM Data for GW190814}\label{sec:combine}

With the release of detailed fits to the GW data for GW190814, including component masses, spin constraints, and inclination \citep{Abbott20}, we can place more meaningful limits on likely merger scenarios and possible electromagnetic counterparts than with the initial NSBH classification.  In particular, the LVC constrained the individual component masses as 2.59$\pm$0.08~$M_{\odot}$ and 23.2$\substack{+1.0\\-0.9}$~$M_{\odot}$ \citep{Abbott20}.  While these final component masses are consistent with the initial NSBH classification, they imply that there is likely very little ejecta mass under most realistic merger models \citep{Faber06,2007NJPh....9...17L,Ferrari10,2011ApJ...736L..21R,Rosswog13}, consistent with the {\tt HasRemnant}=0 statistic.

However, several studies have suggested that the secondary component in this system is likely a NS \citep{Huang20,Zhang20,Zevin20,Tews20}, although see counterarguments that this component may exceed the maximum  NS mass \citep[][]{Akmal98,Heger02,Lattimer07,Foley20}.  Assuming GW190814 resulted from a NSBH system, our limits place constraints both on parameters that are also constrained by the LVC such as the effective inspiral spin parameter (i.e., the total spin with respect to the orbital plane of the binary, $\chi_{\mathrm{eff}}$) and unconstrained parameters such as the NS equation of state.  We consider the effect of varying both of these parameters in the context of our limits on kilonova ejecta mass and velocity.

Using our constraints on ejecta mass and velocity from above for a ``blue'' kilonova, we consider the extent to which we can rule out different values of $\chi_{\mathrm{eff}}$.  For the former, we consider a range of $\chi_{\mathrm{eff}}=0.01$, 0.5, and 0.995 \citep[where the first value is consistent with the effective-one-body approach waveform parameters, which imply $\chi_{\mathrm{eff}}=0.001\substack{+0.059\\-0.056}$, in][]{Abbott20}.  The last value implies a near-maximally spinning BH, which is inconsistent with the general population of BH binaries observed by the LVC \citep[e.g.,][]{Abbott15,Abbott16:GW151226,Abbott17:0814}, but might be obtained in binary populations that evolve from high-mass X-ray binaries \citep[which can have near-maximal spins;][]{McClintock06} or evolve through a common-envelope phase \citep{Livio88,Belczynski16}.

\begin{figure}
    \centering
    \includegraphics[width=0.49\textwidth]{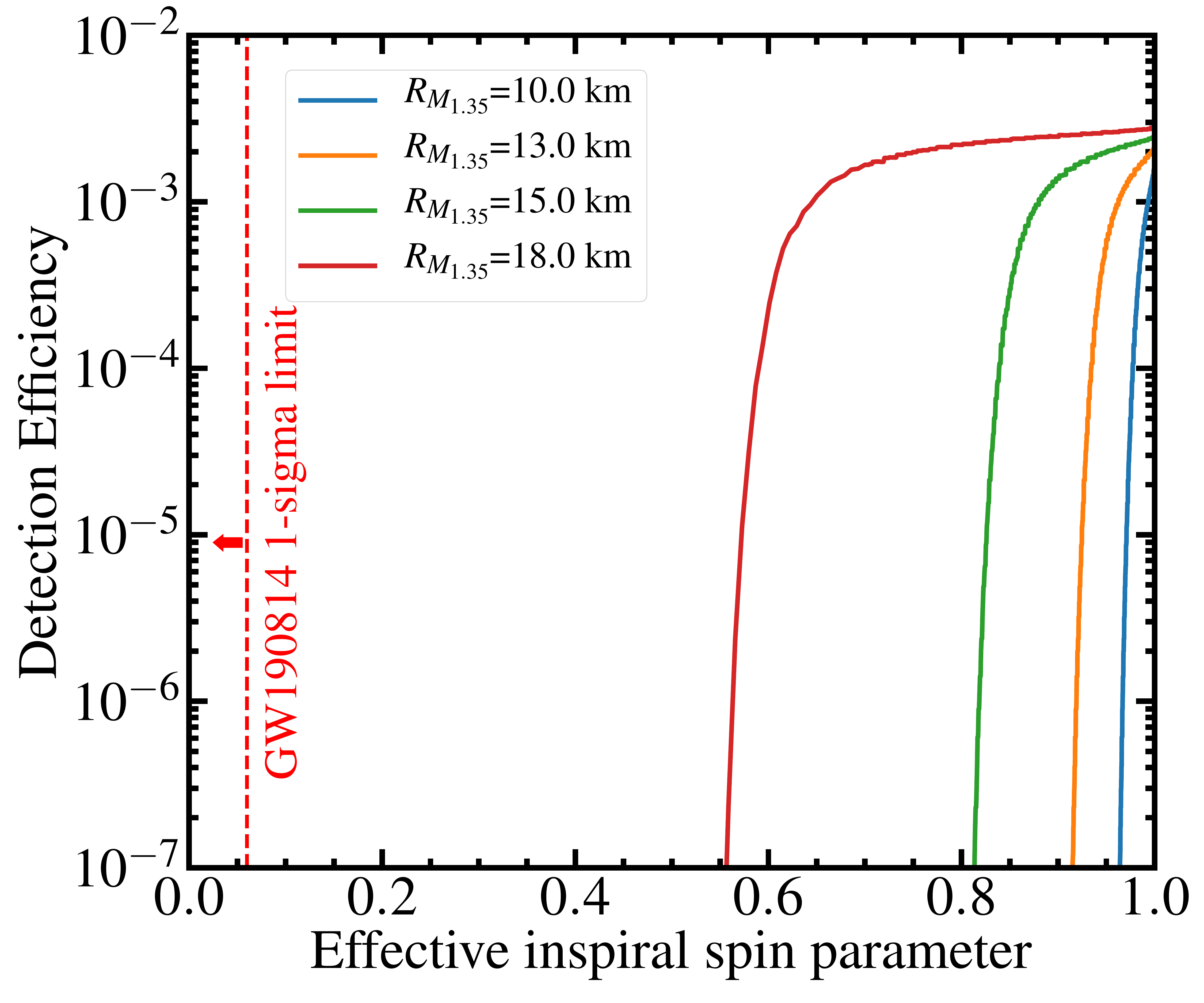}
    \caption{Detection efficiency for kilonovae (described in \autoref{sec:kilonova}) as a function of the effective inspiral parameter \citep[$\chi_{\mathrm{eff}}$, which is constrained to be $<0.06$ for GW190814 by][]{Abbott20}.  We marginalize over the best-fitting component masses for GW190814, which are 23.2$\substack{+0.9\\-1.0}~M_{\odot}$ and 2.59$\pm0.08~M_{\odot}$ based on the effective-one-body approach described by \citet{Abbott20}. We use these parameters to infer kilonova ejecta mass and velocity following methods in \citet{Kawaguchi16} and assuming various equations of state parameterized by radii of a $1.35~M_{\odot}$ NS from 10--18~km.  Based on these constraints, it is not expected that there would be significant ejecta mass with parameters inferred for GW190814 even if the secondary is an NS.}
    \label{fig:gwconstraints}
\end{figure}

Assuming an ejecta mass and velocity as a function of the binary component masses $m_{1}$ and $m_{2}$ and fixed $\chi_{\mathrm{eff}}$, we predict the total ejecta mass and velocity using equations in \citet{Kawaguchi16}.  We then use our blue kilonova models to predict the extent to which we could rule out such a model (\autoref{fig:gwconstraints}, where the color corresponds to $P_{\mathrm{model}}$ as in \autoref{sec:constraints}).  We assume a phenomenological NS equation of state that predicts $R_{\mathrm{NS}}=13.6$~km for a Chandrasekhar-mass NS \citep[H4 in][]{Read09}, whereas a 2.6~$M_{\odot}$ NS would have a radius of $\approx$12~km assuming this equation of state and the formalism in \citet{Kawaguchi16}.  However, given the extreme mass ratio of GW190814, ejecta masses in the range of our blue kilonova limits would require a near maximally spinning BH and an unrealistic equation of state to produce any ejecta mass.

From the group of models, we only rule out the GW190814 system with significance $P_{\mathrm{model}}>20\%$ (see definition of $P_{\mathrm{model}}$ in \autoref{sec:constraints}) when $\chi_{\mathrm{eff}}>0.995$ (\autoref{fig:gwconstraints}).  As we decrease the value of $\chi_{\mathrm{eff}}$, the value of $P_{\mathrm{model}}$ decreases significantly below 0.1\%.

The availability of robust component masses and spin parameters beyond what can be inferred with low-latency LVC parameters (i.e., NSBH classification and {\tt HasRemnant}) has an enormous impact on the predicted observability of EM counterparts.  Furthermore, this problem is unique to NSBH follow-up programs where no plausible counterparts have been found. Specific component masses and spins may or may not yield significant ejecta depending on the assumptions we make above about the specific merger model and equation of state.

\subsection{Constraints on AGN Counterparts to GW190814}\label{sec:agn}

AGNs have been discussed as potential optical counterparts to BBH mergers \citep{Bartos17,2019ApJ...884...22A,Grobner20}, and the discovery of an optical flare timed 34~days after and in the localization region of the BBH merger S190521g/GW190521 \citep{Graham20,Abbott20:GW190521} provides a credible candidate optical counterpart to this phenomenon \citep[although][suggest there is insufficient evidence to confidently associate the AGN with GW190521]{Ashton20}.  AGN activity is a result of massive gas inflows to a supermassive BH, and as a result of that cause compact object mergers have been explored in the literature \citep[e.g.,][]{Stone17}, and so the compact object merger rate is thought to be significantly enhanced in the disks of these systems.  If these systems can form in the disk environment, it is plausible that a significant fraction of LVC counterparts could originate from galaxies with AGNs.

Critically, when a compact object merger occurs in the AGN disk, it can induce an instability in the inflow of gas to the accreting BH, leading to a luminous transient \citep[as is hypothesized to be the case for the S190521g candidate counterpart;][]{McKernan12,2019ApJ...884...22A,Graham20}.  Thus, the AGN model provides some predictive power for candidate optical counterparts, even in cases where no significant mass of baryonic ejecta is expected from the merger itself.

Here we examine what constraints can be placed on the distance to and localization of GW190814 under the assumption that the event occurred in a known AGN inside the GW190814 localization region.  We analyzed the AGN catalog of \citet{Secrest15}, which is selected from Mid-infrared {\it WISE} constraints and contains 1.4 million AGNs down to as faint as $g=26$~mag.  The catalog is estimated to be complete for known AGNs to $>84$\% and for all AGNs with $R<19$~mag.  Therefore, for AGNs with $z<0.1$, the catalog is expected to be close to $>90$\% complete.

For this analysis, we only consider AGNs that (1) are within the 99th percentile localization region of GW190814, and (2) have a redshift that places it within the 99th percentile volume as described in \autoref{sec:candidates}.  Based on these criteria, we do not identify any AGNs with redshifts listed in the \citet{Secrest15} catalog or NED.

However, a major caveat to this analysis is that the majority of AGNs do not have measured redshifts, and so most objects would be ruled out for lying outside the volume we consider here.  For example, we find 1886 AGNs that match only criterion \#1 above.  If we consider what fraction of those objects without spectroscopic redshifts lie within 2\arcsec\ of a source in the PS1-STRM catalog with a photometric redshift that places them in the 99th percentile volume, we are left with only 4 candidates without redshifts in \citet{Secrest15}.  We list these sources along with their photometric redshifts in \autoref{tab:agn}.

%\startlongtable
\begin{deluxetable*}
{lcccccccl}
\tabletypesize{\scriptsize}
\tablecaption{Candidate AGN Hosts for GW190814\label{tab:agn}}
\tablewidth{0pt}
\tablehead{
\colhead{Name\tablenotemark{a}}& 
\colhead{$\alpha$}& 
\colhead{$\delta$}& 
\colhead{Relative Prob.\tablenotemark{b}}&
\colhead{Redshift\tablenotemark{c}} \\
&
(J2000) &
(J2000)
&
&}
\startdata
J012323.69-310826.4 & 01:23:23.70 & $-$31:08:26.43 & 0.476 & 0.021$\pm$0.002  \\
J005203.80-272348.9 & 00:52:03.80 & $-$27:23:48.92 & 0.210 & 0.079$\pm$0.004  \\
J004801.90-215442.2 & 00:48:01.90 & $-$21:54:42.22 & 0.082 & 0.058$\pm$0.001  \\
J004002.26-235053.0 & 00:40:02.27 & $-$23:50:53.03 & 0.232 & 0.071$\pm$0.004  
\enddata
\vspace{0.1in}
\tablenotemark{a}{Name of the AGN in the GW190814 localization region in the WISEA catalog from \citet{Secrest15}.}
\tablenotetext{b}{Relative probability of each AGN given the latest GW190814 map provided by \citet{Abbott20} such that the sum of all probabilities is unity.}
\tablenotetext{c}{Redshift of matching counterpart given in the PS1-STRM catalog, as all candidate AGNs detected in the GW190814 localization region are matched to sources with photometric redshifts in \citet{PS1redshift} apart from WISEA J005720.19-222256.5, whose spectroscopic redshift is derived from \citet{Jones09}.}
\end{deluxetable*}

We note that none of these systems is coincident with any of our candidate counterparts to GW190814 (\autoref{tab:candidates}) or any transients listed in the Transient Name Server to within 5\arcsec.  Thus, while we consider the AGNs as candidate hosts to GW190814, any hypothetical EM counterparts induced by the merger would have to be low luminosity.

The AGN flare luminosity would scale as the total mass of the merging binary $M_{\mathrm{NSBH}}^{2}$ as in \citet{Bartos17} and \citet{2019ApJ...884...22A}.  Following Equation (5) of \citet{Graham20} for the total luminosity of such an EM counterpart with radiative efficiency $\eta=0.1$, kick velocity for the binary in an AGN disk $v_{k}=200$~km~s$^{-1}$, disk gas density $\rho=10^{-10}$~g~cm$^{-3}$, and using the total mass $M_{\rm tot}$ of the GW190814 merger $25.8~M_{\odot}$ \citep{Abbott20}, we find that the luminosity would be $L=1.6\times10^{44}$~erg~s$^{-1}$, or $M_{\rm bol}=-21.8$~mag, which corresponds to 15.1~mag at $D_{L}=241$~Mpc.  A source with this brightness is easily ruled out near our maximal efficiency ($\sim95$\%) assuming it occurred within the 2-week time frame of our observations.  On the other hand, the luminosity is easily scaled down for a lower density in the AGN disk, a higher kick velocity, or a lower radiative efficiency in the AGN.

Based on the associated AGN model of \citet{Graham20}, we would expect to see flaring activity of GW190814 if one of these objects had been associated with the merger.  Therefore, we examined the ASAS-SN Sky Patrol\footnote{\url{https://asas-sn.osu.edu/}} \citep{Shappee14,Kochanek17} to determine whether there is any flaring activity potentially associated with the GW event.  Although variability is detected for WISEA J005203.80-272348.9, J004801.90-215442.2, and J004002.26-235053.0 after the GW190814 merger, none of them exhibits a $>10^{44}$~erg~s$^{-1}$ increase in flux on a timescale of $<8$~weeks from the GW event.  We conclude that these candidates are unlikely to be associated with GW190814.

We also revisit the sources AT~2019nto or AT~2019nup discussed in \autoref{sec:variability}, which we had previously ruled out as being associated with GW190814 owing to significant premerger variability.  This type of variability could occur in the scenario where GW190814 triggered activity in an already active accreting supermassive BH.  In both cases, there is a credible optical counterpart, although both are significantly fainter ($\sim20.8$~mag) than we predict for an AGN counterpart at the distance to GW190814.  Both events have nearly continuous coverage from the ASAS-SN Sky Patrol up to $\sim170$~days from the time of merger.  A source is detected at the site of AT~2019nto twice during that interval $>100$~days from the time of merger, and AT~2019nup is detected almost continuously during that time but with no $>1$~mag increases in brightness.  If either of these objects is associated with an AGN from GW190814, that event would need to occur with an extremely low radiative efficiency ($\eta<0.01$) or from a disk with a much lower gas density ($\rho < 10^{-12}$~g~cm$^{-3}$) compared with the model presented above.

One of the primary difficulties in detecting an AGN counterpart to a GW source is the dynamical timescale for perturbations in the disk to induce enhanced accretion in the AGN and trigger a flare.  This dynamical timescale $t_{\mathrm{dyn}}$ \citep[Equation 6 of][]{Graham20} is only $\sim5$~days assuming that the kick velocity $v_{k}=200$~km~s$^{-1}$.  On the other hand, $t_{\mathrm{dyn}}$ scales as $v_{k}^{-3}$, and so for a marginally lower kick velocity it could easily be outside of the window of our observations (and the flare of significantly lower luminosity, as above).  While this velocity needs to remain small enough that the system remains bound to the disk of the AGN to trigger a flare, scaling the kick velocity in a fixed-mass AGN and at a fixed orbital semimajor axis suggests that $v_{k}\propto M_{\rm tot}^{-1/2}$, or $\sim400$~km~s$^{-1}$ compared with 200~km~s$^{-1}$ for the BBH system of \citet{Graham20}.

Regardless of the presence of an EM counterpart to validate the AGN counterpart hypothesis, if we assume that the event occurred in such an environment we can obtain separate constraints on the distance to this event.  There is a single AGN (WISEA J004506.98-250147.0) that represents $>50\%$ of the normalized two-dimensional probability for GW190814 after renormalizing under the assumption that one of these candidates is the host.  Accounting for the individual uncertainties and weighting by this local two-dimensional probability, we find that the best-fitting redshift is $z=0.052\pm0.017$ or $D_{L}=233\pm80$ Mpc. Compared with the GW190814 distance marginalized over the entire localization region, this inferred distance is more uncertain.

\section{Conclusions}\label{sec:conclusion}

We have presented results from the joint follow-up observations of the LVC NSBH merger GW190814.  Our combined constraints from optical imaging and spectroscopy demonstrate the following.

\begin{enumerate}
    \item There are no plausible optical counterparts detected by our programs or those of any other optical or radio follow-up groups \citep{Gomez19,Dobie19,Andreoni19,Morgan20,Thakur20,Watson20,Vieira20,Ackley20,Alexander21,deWet21}.  Given criteria that we describe in \autoref{sec:candidates}, we are able to rule out all known transient sources detected within the 99th percentile localization of GW190814 and discovered within 2~weeks of the merger time given by the LVC.
    \item Given that there are no plausible counterparts, we are able to rule out kilonovae, GRBs, and SN-like optical counterparts to deep limits using the joint limits from all follow-up observations performed in this study.  While our limits are not constraining in the context of red (low $Y_{e}$) kilonovae, we rule out blue kilonovae ($Y_{e}>0.4$) with $v_{\rm ej}=0.25$c and $M_{\rm ej}>0.3~M_{\odot}$ to 25\% significance.  We also rule out sGRBs similar to those of \citet{Fong15} at $\sim$95\% significance for on-axis events and at $>$50\% significance for viewing angles $\theta_{\rm obs}<17^{\circ}$.  Finally, while our limits can probe luminosities as deep as -16~mag (21~mag at 240~Mpc) across most of the localization region, we cannot rule out events this faint with decline rates comparable to AT~2017gfo at $>$0.3~mag~day$^{-1}$.
    \item Using our joint limits on optical counterparts, we combine our EM follow-up data with the GW data of \citet{Abbott20} to consider scenarios in which NSBH systems would be detectable.  We find that only for near-maximally spinning BHs (where the NSBH system has $\chi_{\mathrm{eff}}>0.995$) can we rule out merger scenarios similar to GW190814 with appreciable ($>$20\%) significance.  While these parameters are inconsistent with those observed from GW190814 (with $\chi_{\mathrm{eff}}<0.06$ for all waveform models), these limits provide a baseline for plausible constraints on future NSBH counterparts.
    \item We consider the possibility that GW190814 occurred in the disk of an AGN similar to the potential counterpart to the BBH merger S190521g proposed by \citet{Graham20}.  Analyzing AGN catalogs, we find only 7 galaxies with AGNs and localized within the volume of GW190814.  This is more than 3 orders of magnitude smaller than the total number of galaxies in the localization region and may provide an efficient search strategy for targeting electromagnetic emission for future compact object mergers.
\end{enumerate}

This analysis was conducted on data collected by three GW follow-up efforts: 1M2H, KAIT, and Las Cumbres, each with independent observational strategies.  Each collaboration used a different set of filters, targeting selection, and timescales that offer a unique constraints on EM counterparts to GW190814.  Our combined data sets enable a more comprehensive and uniform analysis than was immediately possible after the discovery of the event.  The future of GW follow up efforts will benefit from similar analyses using data sharing and communication media such as the Treasure Map \citep{Wyatt20} and the Gamma-ray Coordinates Network \citep{GCN}, open source software and analysis tools such as {\tt teglon} and {\tt gwemlightcurves} \citep{Coughlin20:Hubble}, and increased collaboration within the GW/EM community.

\software{{\tt astropy} \citep{astropy},
          {\tt DoPhot} \citep{Schechter93},
          {\tt hotpants} \citep{Becker15},
          {\tt healpy} \citep{healpy},
          {\tt IDL} \citep{IDL},
          {\tt LCOGTSNpipe} \citep{Valenti16},
          {\tt ligo.skymap} \citep{GTD, GTDSupplement},
          {\tt PypeIt} \citep{pypeit:joss_arXiv, pypeit:zenodo},
          {\tt SExtractor} \citep{sextractor},
          {\tt teglon}}

\facilities{Las Cumbres (SINISTRO), Lick/Shane 3m (Kast), KAIT, Keck:I (MOSFIRE), Keck:II (DEIMOS), Nickel (Direct 2K), SOAR (Goodman), Swope (Direct 4K), Thacher (ACP)}

\begin{acknowledgments}

We thank T.~M.\ Davis, W. Fong, and J.~X.\ Prochaska for helpful comments on this manuscript as well as J.~J.\ Hermes and S.\ Points for support with the SOAR observations.
We thank the staffs at the various observatories where data were obtained for their expert assistance.

Much of this work was performed during the ``Astrophysics in the LIGO/Virgo Era'' meeting at the Aspen Center for Physics during Summer 2019 with C.D.K., D.A.C., I.A., D.O.J., C.R.-B., E.R.-R., A.R., and M.R.S.\ all participating.  The Aspen Center for Physics is supported by National Science Foundation grant PHY-1607611.

The UCSC team is supported in part by NASA grant NNG17PX03C, NSF grant AST-1815935, the Gordon \& Betty Moore Foundation, the Heising-Simons Foundation, and by fellowships from the David and Lucile Packard Foundation to R.J.F.
D.A.C.\ acknowledges support from the National Science Foundation Graduate Research Fellowship under Grant DGE1339067.
A.V.F.’s group at U.C. Berkeley is grateful for financial assistance from the Miller  Institute  for Basic Research in Science (in which A.V.F. is a Miller Senior Fellow), the Christopher  R.\  Redlich  Fund, Steven Nelson, and many other individual donors.
D.E.H.\ was supported by NSF grants PHY-1708081 and PHY-2011997, and the Kavli Institute for Cosmological Physics at the University of Chicago through an endowment from the Kavli Foundation.
Time domain research by D.J.S.\ is supported by NSF grants AST-1821987, 1813466, \& 1908972, and by the Heising-Simons Foundation under grant \#2020-1864.
F.O.E.\ acknowledges support from the FONDECYT grant nr.\ 1201223.
I.A.\ is a CIFAR Azrieli Global Scholar in the Gravity and the Extreme Universe Program and acknowledges support from that program, from the European Research Council (ERC) under the European Union’s Horizon 2020 research and innovation program (grant agreement number 852097), from the Israel Science Foundation (grant number 2752/19), from the United States - Israel Binational Science Foundation (BSF), and from the Israeli Council for Higher Education Alon Fellowship.
J.B.\ is supported by NSF grants AST-1313484 and AST-1911225, as well asby NASA grant 80NSSC19kf1639.
J.C.\ acknowledges support from the Australian Research Council Centre of Excellence for Gravitational Wave Discovery (OzGrav) project number CE170100004.
J.G.B.\ is supported by MINECO project PGC2018-094773-B-C32.
L.S.S.\ acknowledges the financial support from FAPESP through the grant \#2020/03301-5.
M.M.\ is supported by CONICET, CNPq and FAPERJ.
M.R.S.\ is supported by the National Science Foundation Graduate Research Fellowship Program under grant No.\ 184240.
The UM team is supported by NSF grant AST-1910719 and fellowships from the Alfred P.\ Sloan Foundation and the Cottrell Scholar Award to M.S.-S.
N.H. acknowledges support for this work by Israel Science Foundation grant No. 541/17.
R.R.d.C.\ acknowledges the financial support from FAPESP through the grant \#2014/11156-4.
R.R.M.\ acknowledges partial support from project BASAL AFB-$170002$ as well as FONDECYT project N$^{\circ}1170364$.
S.B.R.\ acknowledges support from Conselho Nacional de Desenvolvimento Científico e Tecnológico – CNPq.
T.D.\ is supported by ARC grant FL180100168.
T.L.P.\ acknowledges financial support from CAPES.

This work includes data obtained with the Swope Telescope
at Las Campanas Observatory, Chile, as part of the Swope Time Domain Key Project (PI Piro, Co-Is Burns, Cowperthwaite, Dimitriadis, Drout, Foley, French, Holoien, Hsiao, Kilpatrick, Madore, Phillips, and Rojas-Bravo).
This work makes use of observations from the LCO Network.  The LCO Group is supported by NSF grant AST-1911151.

Some of the data presented herein were obtained at the W. M. Keck Observatory, which is operated as a scientific partnership among the California Institute of Technology, the University of California, and NASA. The Observatory was made possible by the generous financial support of the W. M. Keck Foundation.
The authors wish to recognize and acknowledge the very significant cultural role and reverence that the summit of Maunakea has always had within the indigenous Hawaiian community.  We are most fortunate to have the opportunity to conduct observations from this mountain.

Based in part on observations obtained at the Southern Astrophysical Research (SOAR) telescope, which is a joint project of the Minist\'{e}rio da Ci\^{e}ncia, Tecnologia, Inova\c{c}\~{o}es e Comunica\c{c}\~{o}es (MCTIC) do Brasil, the U.S. National Optical Astronomy Observatory (NOAO), the University of North Carolina at Chapel Hill (UNC), and Michigan State University (MSU).

Research at Lick Observatory is partially supported by a generous gift from Google.
KAIT and its ongoing operation were made possible by donations from Sun Microsystems, Inc., the Hewlett-Packard Company, AutoScope Corporation, Lick Observatory,   the NSF, the University of California, the Sylvia and Jim Katzman Foundation, and the TABASGO Foundation.

\bigskip

\end{acknowledgments}

\bibliography{gw190814}

\appendix

\startlongtable
\begin{deluxetable}
{lcccccc}
\tabletypesize{\scriptsize}
\tablecaption{Optical/IR Imaging of the GW190814 Localization Region\label{tab:observations}}
\tablewidth{0pt}
\tablehead{
\colhead{Source\tablenotemark{\footnotesize a}} & 
\colhead{$\alpha$} & 
\colhead{$\delta$} & 
\colhead{Date} &
\colhead{Filter} &
\colhead{Magnitude Limit\tablenotemark{\footnotesize b}} \\
&
(J2000) &
(J2000) &
(MJD) &
&
(3$\sigma$)
}
\startdata
S & 01:27:55.560 & -34:41:50.280 & 58715.2170 & $r$ & 20.64 \\
S & 01:27:56.496 & -34:42:01.800 & 58716.3976 & $r$ & 21.04 \\
S & 01:27:58.248 & -34:42:09.360 & 58718.3350 & $r$ & 21.12 \\
S & 01:27:46.752 & -34:42:55.080 & 58736.1369 & $r$ & 20.69 \\
S & 01:32:06.792 & -34:12:16.560 & 58713.2037 & $r$ & 20.37 \\
S & 01:32:07.752 & -34:13:05.520 & 58714.2115 & $r$ & 19.58 \\
S & 01:32:09.456 & -34:11:59.640 & 58715.3197 & $r$ & 21.03 \\
S & 01:32:12.288 & -34:12:12.240 & 58716.2327 & $r$ & 20.73 \\
\enddata
%\tablecomments{}
\tablenotetext{a}{Surveys correspond to KAIT (K), Keck/MOSFIRE (M), Las Cumbres Observatory (L), Nickel (N), Swope (S), and Thacher (T) imaging as described in \autoref{sec:obs}.}
\tablenotetext{b}{In-band 3$\sigma$ limit for the reported image as described in \autoref{sec:obs} and \autoref{sec:limits}.  The ``Clear'' filter is abbreviated as ``C'' for all KAIT observations.  All magnitudes are on the AB system \citep{Oke83}.}
\end{deluxetable}

\startlongtable
\begin{deluxetable*}
{lccccccl}
\tabletypesize{\scriptsize}
\tablecaption{Candidate Electromagnetic Counterparts to GW190814\label{tab:candidates}}
\tablewidth{0pt}
\tablehead{
\colhead{Name}& 
\colhead{$\alpha$}& 
\colhead{$\delta$}& 
\colhead{Cumulative LVC Prob.\tablenotemark{\footnotesize  a}}&
\colhead{Discovery Date}&
\colhead{Redshift\tablenotemark{\footnotesize  b}}&
\colhead{Absolute Magnitude\tablenotemark{\footnotesize  c}}& 
\colhead{Note\tablenotemark{\footnotesize d}}\\
&
(J2000) &
(J2000) &
&
(MJD) &
&
(AB mag) &
}
\startdata
2019nmd       & 00:51:29.004 & -22:28:16.96 & 0.9626 & 58710.27800 & --                         & --                              & (MP) 2010 GA33                     \\
2019nme       & 00:50:32.558 & -22:13:33.70 & 0.9723 & 58710.27800 & --                         & --                              & (MP) 2005 OL33                     \\
2019noq       & 00:48:47.882 & -25:18:23.46 & 0.3030 & 58710.58500 & 0.05907$\pm$0.00033 (s)    & $i$=$-$17.28$\pm$0.25 (0.7 d)  & (SN) II                            \\
2019nor       & 00:49:51.992 & -24:16:17.71 & 0.1808 & 58710.58500 & 0.43652$\pm$0.00015 (s)    & $r$=$-$22.33$\pm$0.23 (0.7 d)  & (Z)                                \\
2019npd       & 00:46:56.705 & -25:22:36.66 & 0.5377 & 58710.28000 & 0.00081$\pm$0.00001 (s)    & $i$=$-$8.68$\pm$0.30 (0.4 d)   & (Z; Sculptor Galaxy)                                \\
2019npe       & 00:41:33.330 & -23:44:31.94 & 0.8787 & 58710.27600 & 0.17233$\pm$0.02355 (PS1)  & $z$=$-$19.55$\pm$0.30 (0.4 d)  & (Z)                                \\
2019npf       & 00:44:25.379 & -23:11:52.08 & 0.7192 & 58710.27600 & 0.17953$\pm$0.02702 (PS1)  & $z$=$-$20.28$\pm$0.30 (0.4 d)  & (Z)                                \\
2019nph       & 00:58:07.613 & -27:12:20.27 & 0.9362 & 58710.30899 & 0.23849$\pm$0.06683 (PS1)  & $r$=$-$20.88$\pm$0.30 (0.4 d)  & (Z)                                \\
\enddata
%\tablecomments{}
\tablenotetext{a}{Cumulative probability within the {\tt GW190814\_skymap} map provided by \citet{Abbott20}.}
\tablenotetext{b}{Redshifts are identified as spectroscopic (s) or photometric (by source) as described in \autoref{sec:candidates}.}
\tablenotetext{c}{We indicate absolute magnitude at the earliest detection epoch (given in parentheses in observer-frame days relative to the GW190814 merger).}
\tablenotetext{d}{We rule out each source based on classification as likely minor planets (MP), supernovae (SN), pre-merger variability (VAR), a redshift inconsistent with the GW190814 volume (Z), or photometric evolution (PHOT) as described in \autoref{sec:candidates}.  For minor planets we give the identified minor planet, and for supernovae we give the type described in \autoref{sec:candspec}.}
\tablenotetext{e}{The photometry used to rule out 2019qbz is presented in \citet{Ackley20}.  2019qbz has a relatively flat light curve across $\approx$6~days of follow up as shown in \autoref{fig:cand-lightcurves}.}
\end{deluxetable*}

\startlongtable
\begin{deluxetable}
{lccccc}
\tabletypesize{\scriptsize}
\tablecaption{Photometry of Candidate Electromagnetic Counterparts to GW190814\label{tab:cand-photometry}}
\tablewidth{0pt}
\tablehead{
\colhead{Name} & 
\colhead{MJD} &
\colhead{Filter} &
\colhead{$m$} &
\colhead{$\sigma_{m}$} &
\colhead{Source}\\
&
&
(mag) &
(mag) &
}
\startdata
2019noq & 58713.2883 & $r$& 20.12 & 0.10 & Swope \\
2019noq & 58714.2699 & $r$& 20.39 & 0.10 & Swope \\
2019noq & 58715.2022 & $r$& 20.44 & 0.09 & Swope \\
2019noq & 58716.3051 & $r$& 20.46 & 0.06 & Swope \\
2019noq & 58720.3270 & $r$& 20.42 & 0.04 & Swope \\
2019noq & 58725.2115 & $r$& 20.42 & 0.04 & Swope \\
2019noq & 58737.3559 & $r$& 20.57 & 0.08 & Swope \\
2019npd & 58723.4640 & $r$& 19.67 & 0.11 & Thacher \\
\enddata
\tablecomments{All photometry of candidate counterparts (\autoref{tab:candidates}) to GW190814 from follow up and search observations presented in this paper.  We use these data along with photometry presented in \citet{Andreoni19}, \citet{Ackley20}, and \citet{Morgan20} to classify all candidates.}
\end{deluxetable}

\end{document}